\documentclass[10pt,final,journal]{IEEEtran}
\usepackage[switch]{lineno} 
\usepackage{tabularx} 
\usepackage{booktabs}
\usepackage{multirow}
\usepackage{makecell}

\usepackage{stmaryrd}
\usepackage{graphicx}
\usepackage{amsthm} 
\usepackage{amsmath} 
\usepackage{bm}
\usepackage{graphicx}
\usepackage{epstopdf}
\usepackage{color}
\usepackage{float}
\usepackage[colorlinks=false,linkcolor=black,urlcolor=black,bookmarksopen=true, hidelinks]{hyperref}
\usepackage{bookmark}
\usepackage[flushleft]{threeparttable}
\usepackage{stackengine}
\usepackage{scalerel}
\usepackage{array}

\usepackage{xcolor,soul}

\usepackage{cite}

\ifCLASSINFOpdf
\else
\fi
\usepackage{amsfonts,amssymb}

\usepackage{amsmath}
\usepackage{algorithm}
\usepackage{algpseudocode}
\ifCLASSOPTIONcompsoc
 \usepackage[caption=false,font=normalsize,labelfont=sf,textfont=sf]{subfig}
\else
 \usepackage[caption=false,font=footnotesize]{subfig}
\fi
\begin{document}
%

\title{A New TOA Localization and Synchronization System with Virtually Synchronized Periodic Asymmetric Ranging Network}

\author{Sihao~Zhao, 
		Xiao-Ping~Zhang, \textit{Fellow, IEEE,}
        Xiaowei~Cui,
        and~Mingquan~Lu
\thanks{This work was supported in part by the Natural Sciences and Engineering Research Council of Canada (NSERC), Grant No. RGPIN-2020-04661. \textit{(Corresponding author: Xiao-Ping Zhang)}}
\thanks{S. Zhao, X.-P. Zhang are with the Department of Electrical, Computer and Biomedical Engineering, Ryerson University, Toronto, ON M5B 2K3, Canada (e-mail: sihao.zhao@ryerson.ca; xzhang@ryerson.ca).}
\thanks{X. Cui is with the Department of Electronic Engineering,
	Tsinghua University, Beijing 100084, China (e-mail: cxw2005@tsinghua.edu.cn).}
\thanks{M. Lu is with the Department of Electronic Engineering,
	Beijing National Research Center for Information Science and Technology, Tsinghua University, Beijing 100084, China. (e-mail: lumq@tsinghua.edu.cn).}
\thanks{Copyright (c) 20xx IEEE. Personal use of this material is permitted. However, permission to use this material for any other purposes must be obtained from the IEEE by sending a request to pubs-permissions@ieee.org.}
}

\markboth{}%
{Shell \MakeLowercase{\textit{et al.}}: Bare Demo of IEEEtran.cls for IEEE Journals}
%




\maketitle

\begin{abstract}

In this article, we design a new time-of-arrival (TOA) system for simultaneous user device (UD) localization and synchronization with a periodic asymmetric ranging network, namely PARN. The PARN includes one primary anchor node (PAN) transmitting and receiving signals, and many secondary ANs (SAN) only receiving signals. All the UDs can transmit and receive signals. The PAN periodically transmits sync signal and the UD transmits response signal after reception of the sync signal. Using TOA measurements from the periodic sync signal at SANs, we develop a Kalman filtering method to virtually synchronize ANs with high accuracy estimation of clock parameters. Employing the virtual synchronization, and TOA measurements from the response signal and sync signal, we then develop a maximum likelihood (ML) approach, namely ML-LAS, to simultaneously localize and synchronize a moving UD. We analyze the UD localization and synchronization error, and derive the Cram\'er-Rao lower bound (CRLB). Different from existing asymmetric ranging network-based TOA systems, the new PARN i) uses the periodic sync signals at the SAN to exploit the temporal correlated clock information for high accuracy virtual synchronization, and ii) compensates the UD movement and clock drift using various TOA measurements to achieve consistent and simultaneous localization and synchronization performance. Numerical results verify the theoretical analysis that the new system has high accuracy in AN clock offset estimation and simultaneous localization and synchronization for a moving UD. We implement a prototype hardware system and demonstrate the feasibility and superiority of the PARN in real-world applications by experiments.

\end{abstract}



%
\IEEEpeerreviewmaketitle

\section{Introduction}\label{Introduction}
%
%
%
%
\IEEEPARstart{L}{ocation} information plays a key role in a variety of applications relating to our daily life and is drawing more attention from academic and industrial communities. A typical wireless localization system usually comprises anchor nodes (ANs) at known positions and user devices (UDs) that need to be localized and sometimes synchronized. For any localization technique, some kinds of measurements between the UD and ANs need to be obtained to achieve localization and synchronization. Such measurements include time-of-arrival (TOA), direction-of-arrival (DOA), received signal strength (RSS) \cite{shao2014efficient,li2013smartphone,luo2019novel,wang2012novel,zhao2020closed,hu2017robust,shi2019anchor}. The TOA-based technique has high accuracy and is adopted in numerous modern localization systems and applications such as global navigation satellite system (GNSS), Radar, acoustic/ultra-sound based tracking, ultra-wide band (UWB) indoor localization, wireless sensor networks (WSN) and Internet of Things (IoT) localization \cite{kuutti2018survey,hofmann2007gnss,xue2019locate,conti2019soft,an2020distributed,zafari2019survey}.


TOA-based systems mainly fall into two different categories, namely one-way and two-way TOA \cite{mccrady2000mobile,shi2020sequential}, respectively, according to how many communications are conducted between a UD and an AN. One-way TOA is obtained by measuring the one-way signal transmission and reception time based on the respective clock sources of the AN and UD. It requires synchronization between ANs. There are three ways to achieve such synchronization, i.e., high accuracy atomic clock and control stations in GNSS \cite{kaplan2005understanding,zhao2019single,misra2006global,chen2016space}, cable connection in regional applications \cite{zhang2009real,segura2012ultra,zwirello2014realization}, and pair-wise wireless communications between ANs \cite{hamer2018self,shi2019blas,shi2020sequential,wang2020tdoa}. However, high accuracy clock sources are usually atomic clocks such as cesium and rubidium clocks, which are expensive and cumbersome \cite{zhang2019ulpac}, and thus are not suitable for consumer level devices such as IoT systems. Cable connection is not flexible for deployment and is difficult to maintain. Pair-wise wireless communications between ANs require more temporal and/or spectral expenses and thus puts a limit on the number of ANs and lower the scalability of the system.

For two-way TOA, measurements are obtained from round-trip communications between the AN and the UD. Compared with one-way TOA, the two-way TOA scheme has the advantage that no synchronization between ANs are required and thus is easy to implement \cite{gholami2016tw,lazzari2017numerical,zou2017joint,nevat2016location,bialer2016two,wu2019coordinate}. However, two-way TOA requires both ANs and UDs to be a transceiver. This adds the cost of the system and is less power efficient. Furthermore, two-way communication has heavier air traffic than one-way communication, resulting in low efficiency for scenarios with dense UDs and/or ANs.

A variation of the TOA-based system, referred to as asymmetric ranging \cite{wang2011robust,chepuri2012joint} or elliptical localization \cite{zhou2010indoor,rui2014elliptic,al2020elliptic,rui2015efficient,amiri2019efficient,xu2012high}, does not require expensive synchronization as in one-way TOA or dense communication as in two-way TOA. The basic idea is that one AN transmits signal (it may be able to receive as well), the UD reflects/relays it or transmits another signal after reception, and the other ANs receive signals from both the transmitter AN and the UD. Due to the asymmetric feature that one AN transmits (may also receive) signal and other ANs only receive signal, we call this scheme ``asymmetric ranging network'' in this article. In such a system, the position and clock offset of a stationary UD are determined using the TOA measurements obtained from the communications both between the transmitting AN and the receiving ANs and between the ANs and the UD. For all the ANs, there is only a single one-way communication from the transmitting AN to all the other receiving ANs. For the UD, there is only one signal reception and one signal transmission. Therefore, compared with one-way TOA, it does not require expensive clocks, cable connection or pair-wise communication between ANs. The communication expense between the UDs and ANs is reduced compared with two-way TOA. In addition, this network has better scalability because the number of the receiving ANs can be increased infinitely. The cost and power consumption of the system can be lowered because the receiver ANs do not have to transmit signals.

However, localization systems incorporating the asymmetric ranging network have the following limitations. i) The temporal correlated AN clock parameters, which partially determine the UD localization accuracy, are estimated only using current or a few recent TOA measurements, and thus have limited accuracy. ii) The assumption of stationary UD does not apply to localization and synchronization for a moving UD, and will result in uncompensated error.

In this article, we design a new TOA localization and synchronization system based on a periodic asymmetric ranging network, namely PARN. In this system, incorporating the asymmetric ranging principle, one primary AN (PAN) periodically transmits the \textit{sync} signal and all the secondary ANs (SAN) and the UD receive it. The UD then transmits the \textit{response} signal after a known delay and all the ANs receive it. Different from other systems based on the asymmetric ranging network, the PAN in the new PARN periodically transmits the sync signal, which is received by each SAN to periodically form multiple TOA measurements. These periodical TOA measurements contain more temporal clock information than a single TOA measurement. High accuracy of network synchronization and UD positioning and timing can be achieved if this abundant temporal clock information is exploited. What's more, we model the motion of the UD with a constant velocity during a short period, and utilize TOA measurements from the response signal formed by the ANs and from the sync signal received by the UD in the PARN along with movement and clock drift compensation. This enables the simultaneous estimation of position and clock offset for a moving UD.

To achieve high-accuracy virtual synchronization for the ANs in the new PARN, we first propose a distributed Kalman filtering method that fully utilizes historical TOA measurements from the periodic PAN signal. The initialization of the filter and the benefit of the AN virtual synchronization to the UD localization and synchronization are presented as a guidance for practical use. Then, to achieve simultaneous localization and synchronization for a moving UD, we propose a maximum likelihood (ML)-based method, namely ML-LAS, which employs TOA measurements from the sync signal and the response signal along with the virtual synchronization results. We conduct error analysis and derive the Cram\'er-Rao lower bound (CRLB) for the ML-LAS method. We also derive the localization and synchronization error with deviated UD velocity and clock drift as a theoretical guidance for real applications.


Numerical simulations of the new PARN are conducted. Results verify the theoretical analysis that the system achieves i) virtual synchronization of the ANs with high accuracy by the proposed Kalman filter-based method, which utilizes historical TOA measurements, and ii) optimal localization and synchronization for a moving UD by the proposed ML-LAS method. We implement a prototype of the proposed PARN using consumer-level embedded hardware. Real-world experiments are conducted to evaluate its performance. Results validate the performance of the virtual synchronization method and the proposed ML-LAS method, all consistent with theoretical analysis, demonstrating that the proposed new system and methods are feasible for real-world applications.


The rest of the article is organized as follows. In Section II, the design of PARN including its measurement and clock model is presented and the localization and synchronization problem based on TOA measurements is formulated. A Kalman filter-based method for virtual synchronization of the SANs is proposed in Section III. The optimal localization and synchronization method for the PARN, namely ML-LAS, and its iterative algorithm are presented in detail in Section IV. The estimation performance of the proposed ML-LAS method is analyzed in Section V. Simulations are conducted to evaluate the performance of the Kalman filter-based virtual synchronization for SANs and the proposed ML-LAS method for the UD in different cases in Section VI. Section VII presents the real-world experimental results using the prototype system based on the proposed PARN scheme. Section VIII concludes this article.
 
Main notations are summarized in Table \ref{table_notation}.

\begin{table}[!t]
	\caption{Notation List}
	\label{table_notation}
	\centering
	\begin{tabular}{l p{5.5cm}}
		\toprule
		lowercase $x$&  scalar\\
		bold lowercase $\boldsymbol{x}$ & vector\\
		bold uppercase $\bm{X}$ & matrix\\
		$\Vert \boldsymbol{x} \Vert$ & Euclidean norm of a vector\\
		$\Vert \boldsymbol{x}\Vert _{\bm{W}}^2$ & square of Mahalanobis norm, i.e., $\boldsymbol{x}^T\bm{W}\boldsymbol{x}$\\
		$i$, $j$ & indices of variables\\
		$[\boldsymbol{x}]_{i}$ &the $i$-th element of a vector\\
		$\mathrm{tr}(\bm{X})$ & trace of a matrix\\
		$[\bm{X}]_{i,:}$, $[\bm{X}]_{:,j}$ &the $i$-th row and the $j$-th column of a matrix, respectively\\
		$[\bm{X}]_{i,j}$ &entry at the $i$-th row and the $j$-th column of a matrix\\
		$|\bm{X}|$ &determinant of a matrix\\
		$\mathbb{E}[\cdot]$ & expectation operator \\
		$\mathrm{diag}(\cdot)$ & diagonal matrix with the elements inside\\
		$M$ & number of ANs\\
		$N$ & dimension of all the position and velocity vectors, i.e., $N=2$ in 2D case and $N=3$ in 3D case\\
		$\boldsymbol{0}_{M}$ & $M$-element vector with all-zero elements\\
		$\boldsymbol{p}_{i}$ & position vector of AN \#$i$\\
		$\boldsymbol{p}_u$ &  unknown position vector of UD\\
		$\boldsymbol{v}_u$ &  velocity vector of UD\\
		$\boldsymbol{e},\boldsymbol{l}$ & unit line-of-sight (LOS) vector from the UD to the AN at the UD transmission and reception time, respectively\\
		$t$ & actual time based-on the reference clock\\
		$\delta t$ & delay interval from signal reception to signal transmission of the UD\\
		$b$, $\omega$ &  clock offset and clock drift with respect to the actual time\\
		$c$ & propagation speed of the signal\\
		$\rho_{i}$ & response-TOA measurement of AN \#$i$ upon reception of the response signal from the UD\\
		$\tau_{i}$,$\tau_u$ & sync-TOA measurement of AN \#$i$ and UD, respectively, upon reception of the sync signal from the primary AN (AN \#1)\\
		$d_{ij}$ &  physical distance between AN \#$i$ and AN \#$j$\\
		$\boldsymbol{\theta}$ &  parameter vector\\
		$\varepsilon$, $\sigma^2$ &  Gaussian random error and variance\\
		$\mathcal{F}$ &  Fisher information matrix\\
		$\bm{W}$ & weighting matrix\\
		$\bm{G}$ & design matrix\\
		$\mu$& estimation bias\\
		$\bm{Q}$ & process noise covariance matrix\\
		$\bm{P}$ & estimation error covariance matrix\\
		$\bm{\Phi}$ & state transition matrix\\
		\bottomrule
	\end{tabular}
\end{table}

\section{System Design} \label{problem}
\subsection{Localization and Synchronization System based on Periodic Asymmetric Ranging Network}
\begin{figure}
	\centering
	\includegraphics[width=1\linewidth]{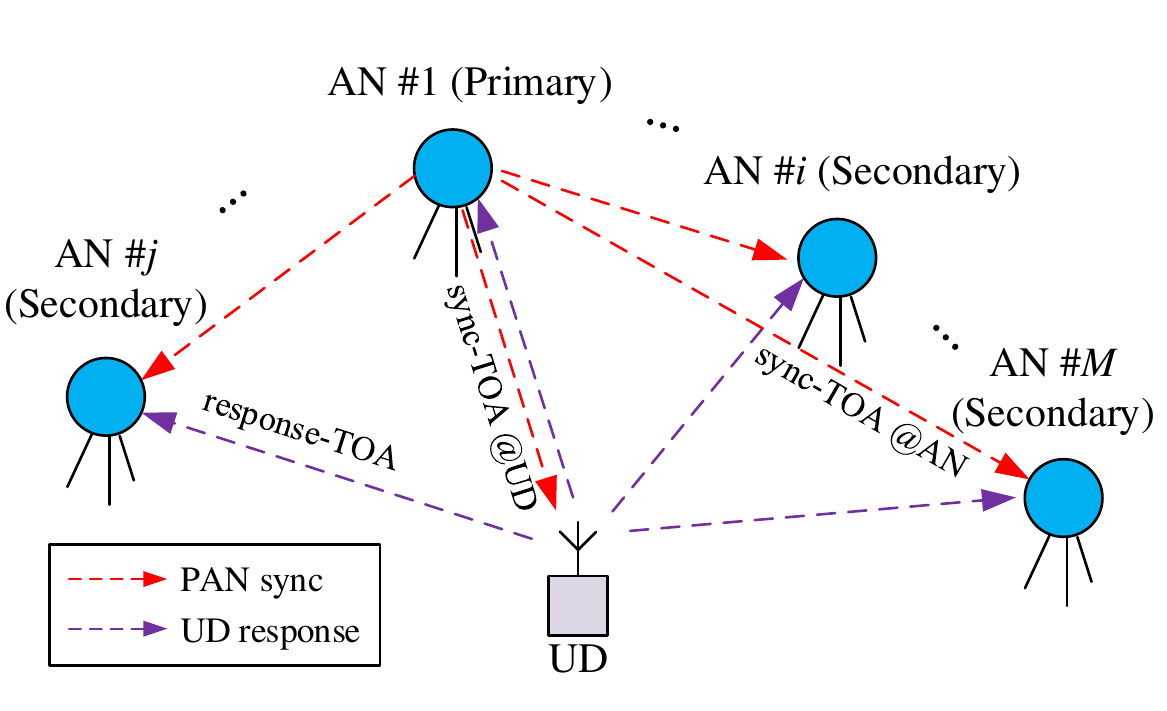}
	\caption{Localization and synchronization system based on periodic asymmetric ranging network (PARN). In one period, AN \#1 (PAN) transmits sync signal and other ANs (SAN) and the UD receive. Then UD transmits response signal after a delay from reception of sync, and SANs receive. Different UDs have different delays so that the system can contain multiple UDs. Sync-TOA measurements are obtained at both the UD and SAN ends, and response-TOA measurements are formed at the AN ends.
	}
	\label{fig:systemfig}
\end{figure}

We design a periodic asymmetric ranging-based localization and synchronization system, namely PARN, as shown in Fig. \ref{fig:systemfig}. In this system, there are $M$ ANs placed at known positions. We denote the coordinate of the $i$-th AN by $\boldsymbol{p}_i$, where $i=1,\cdots,M$. Without loss of generality, AN \#1 is assigned as the primary AN (PAN) that both transmits and receives signals. The other $(M-1)$ ANs only receive signals from both the PAN and the UD, and are named as secondary AN (SAN). The ANs are all asynchronous, i.e., their internal clocks run independently. The position and clock offset of the UD are unknowns to be determined. The positions of the ANs $\boldsymbol{p}_i$ and the position of the UD, denoted by $\boldsymbol{p}_u$, are of $N$ dimension ($N=2$ for 2D cases and $N=3$ for 3D cases), i.e., $\boldsymbol{p}_{i} \text{, } \boldsymbol{p}_{u} \in \mathbb{R}^{N}$.

As shown in Fig. \ref{fig:systemfig}, when the PARN is in operation, AN \#1 transmits sync signal periodically and all SANs and the UDs receive this signal. In one period of PAN transmission, TOA measurements upon reception of the PAN sync signal, namely sync-TOA, are formed at the SAN end and the UD end. Then, after a delay, the UD transmits response signal and all the ANs receive to form TOA measurements, namely response-TOA. Different UDs have different response delays such that multiple UDs can work in one period without collisions. Only the PAN and UDs in this system receive and transmit signals, leading to a reduced communication expense. In addition, the SANs only passively receive signals, and do not need to equip transmission modules. This will cut the cost of the whole system and make the number of SANs expandable. 

\begin{figure}
	\centering
	\includegraphics[width=1\linewidth]{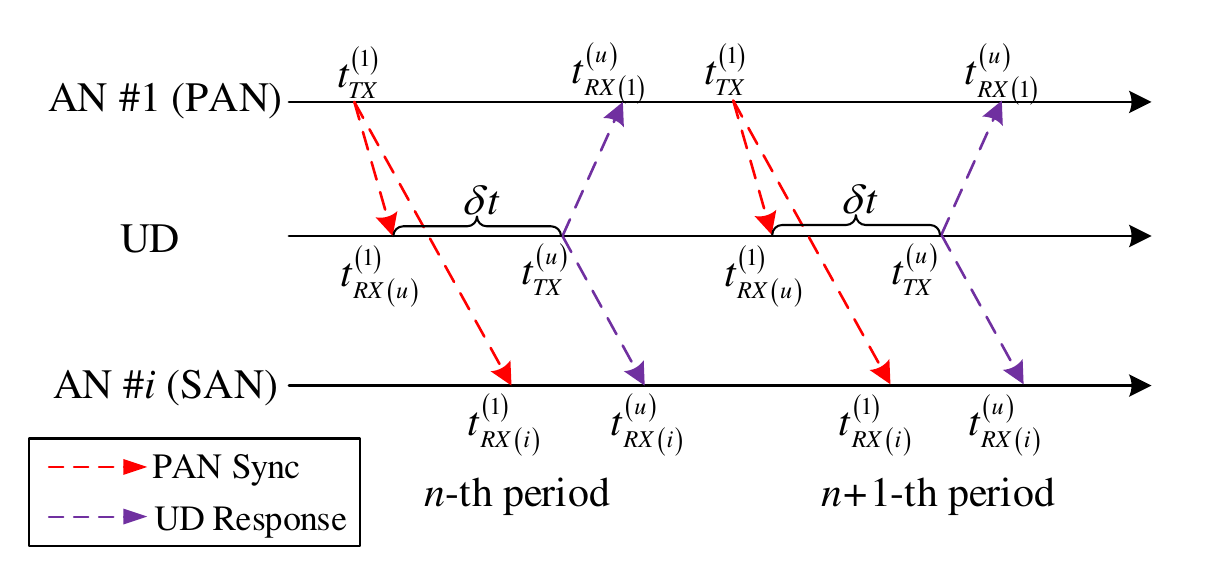}
	\caption{Timing diagram of the PARN communication protocol. One communication period includes PAN transmission, UD reception, SAN reception, UD transmission and AN reception.
	}
	\label{fig:timediagram}
\end{figure}

The timing diagram of the communication in the PARN is depicted in Fig. \ref{fig:timediagram}, where the lower-case ``$t$'' represents the actual time based-on the reference clock, $t_{TX}^{(1)}$ is the sync signal transmission time, the superscript ``$(1)$'' represents that the signal is transmitted by AN \#1, $t_{RX(u)}^{(1)}$ and $t_{RX(i)}^{(1)}$ are the reception times of the sync signal at the UD and AN \#$i$, respectively, the subscript ``$RX(u)$'' and ``$RX(i)$'' represents the reception at the UD and AN \#$i$, respectively, $\delta t$ is the delay interval from the sync signal reception to the response signal transmission, $t_{TX}^{(u)}$ is the transmission time of the response signal, $t_{RX(1)}^{(u)}$ and $t_{RX(i)}^{(u)}$ are the reception times of the response signal at AN \#1 and AN \#$i$, respectively.

With multiple periods of such wireless communications, a series of sync-TOA measurements are collected by each SAN. This enables the utilization of all historical measurements to achieve accurate estimation of the clock parameters, as will be shown in Section \ref{virtualsync}. In one period of communication, one sync-TOA measurement is obtained at the UD end and $M$ response-TOAs are obtained by all the ANs. With these TOA measurements and the estimated SAN clock parameters collected at a central computation unit, we are able to determine the UD position and clock offset simultaneously as will be shown in Section \ref{locmethod}.


\subsection{Clock and UD Movement Models}\label{clockandmovemodel}
Both the AN and UD have its own clock source running independently. They have clock offset and drift with respect to a reference clock. We denote the clock offset by $b_i$ for AN \#$i$ and $b_u$ for the UD, respectively, and the clock drift by $\omega_i$ for AN \#$i$ and $\omega_u$ for the UD, respectively.


In most cases, it is not necessary to have a perfect reference clock, and a local reference clock is sufficient. Without loss of generality, we set the clock of AN \#1 as the reference clock, i.e., $b_1(t)=0 \text{, }\omega_1(t)=0$, where $t$ is the time from the reference clock. 

We first take the clock parameters of the AN as an example. Note that both the clock offset $b_i$ and the clock drift $\omega_i$ vary with the actual time. Following the random walk model given by \cite{axelrad1996gps,brown2012introduction}, we model the clock parameters as
\begin{equation} \label{eq:clockbomega}
\left[\begin{matrix}
b_i(n)\\
\omega_i(n)
\end{matrix}\right]=\bm{\Phi}
\left[
\begin{matrix}
b_i(n-1)\\
\omega_i(n-1)
\end{matrix}
\right]+\boldsymbol{\eta}_i(n-1)
\text{, } i=2,\cdots,M,
\end{equation}
where $n$ is the index of the discrete time, $\bm{\Phi}$ is the state transition matrix as given by
$
\bm{\Phi}=\left[
\begin{matrix}
1 & \Delta t\\
0 & 1
\end{matrix}
\right] \text{,}
$
$\Delta t$ is the time interval, and the noise
$
\boldsymbol{\eta}_i=\left[
\begin{matrix}
\eta_{b_i} & \eta_{\omega_i}
\end{matrix}
\right]^T \text{,}
$
with both $\eta_{b_i}$ and $\eta_{\omega_i}$ being independent zero-mean white noises. Its covariance is
\begin{equation} \label{eq:Qclock}
\bm{Q}=\left[
\begin{matrix}
s_b \Delta t+s_{\omega} \frac{\Delta t^3}{3}& s_{\omega} \frac{\Delta t^2}{2}\\
s_{\omega} \frac{\Delta t^2}{2}& s_{\omega} \Delta t
\end{matrix}
\right] \text{,}
\end{equation}
where $s_b$ and $s_{\omega}$ are the spectral amplitudes of the clock offset and drift, respectively.


The clock offset and drift of the UD have the same model as (\ref{eq:clockbomega}). For the UD movement, we assume the UD velocity is constant during a short time period. Thus, the UD position at different instants $t_1$ and $t_2$ has a relation as given by
\begin{equation} \label{eq:posvel}
\boldsymbol{p}_u(t_2) = \boldsymbol{p}_u(t_1) + \boldsymbol{v}_u \cdot (t_2-t_1) \text{.}
\end{equation}

\subsection{TOA Measurement Model}

\subsubsection{TOA for PAN sync Signal (sync-TOA)}

The PAN transmits the sync signal and the SANs and the UD receive to form sync-TOA measurements. For AN \#$i$ ($i=2,\cdots,M$), the sync-TOA measurement, denoted by $\tau_i$, is the difference of the local reception time at AN \# $i$ and the local PAN transmission time plus measurement noise, i.e.,
\begin{align} \label{eq:tauANiAwo0}
	\tau_i = \left(t_{RX(i)}^{(1)}+b_i\left(t_{RX(i)}^{(1)}\right)\right) -t_{TX}^{(1)} +\varepsilon_{i}  \text{, } i=2,\cdots,M,
\end{align}
where $t_{RX(i)}^{(1)}$ is AN \#$i$'s reception time of the sync signal from AN \#1, the subscript ``$RX(i)$'' represents the reception at AN \#$i$, the superscript ``$(1)$'' represents that the signal is transmitted by AN \#1, $t_{TX}^{(1)}$ is the sync signal transmission time, the superscript ``$(1)$'' also represents that the signal is transmitted by AN \#1, and $\varepsilon_{i}$ is the measurement noise, following independent zero mean Gaussian distribution with a variance of $\sigma_{i}^2$, i.e., $\varepsilon_{i} \sim \mathcal{N}(0,\sigma_{i}^2)$.

We denote the physical distance between AN \#1 and AN \#$i$ by $d_{i1}$, and $d_{i1}=\Vert \boldsymbol{p}_i - \boldsymbol{p}_1\Vert$. We note that this distance equals to the time difference of the signal reception and transmission multiplied by the signal propagation speed $c$, i.e., $d_{i1}=c\left(t_{RX(i)}^{(1)}-t_{TX}^{(1)}\right)$.
Then, we re-write the sync-TOA measurement as the sum of the true signal propagation time and the clock offset. Thus, we come to
\begin{align} \label{eq:tauANiAwo1}
\tau_i = \frac{d_{i1}}{c}+ b_i\left(t_{RX(i)}^{(1)}\right) + \varepsilon_{i} \text{, } i=2,\cdots,M,
\end{align}

Similar to (\ref{eq:tauANiAwo1}), we derive the sync-TOA measurement at the UD upon reception of the sync signal as
\begin{align} \label{eq:tauUD}
\tau_u = \frac{\left\Vert\boldsymbol{p}_1-\boldsymbol{p}_u\left(t_{RX(u)}^{(1)}\right)\right\Vert}{c}+ b_u\left(t_{RX(i)}^{(1)}\right) + \varepsilon_{u} \text{,}
\end{align}
where $t_{RX(u)}^{(1)}$ is the time at the UD upon reception of sync signal from AN \#1, the superscript ``(1)'' represents that the signal is from AN \#1, the subscript ``$RX(u)$'' represents reception by the UD, $b_u$ is the UD clock offset with respect to AN \#1, $\varepsilon_{u}$ is the measurement noise, following a zero-mean Gaussian distribution with a variance of $\sigma_{u}^2$, i.e., $\varepsilon_{u} \sim \mathcal{N}(0,\sigma_{u}^2)$. 

\subsubsection{TOA for UD response Signal (response-TOA)}
The UD transmits the response signal after a delay from the reception of the sync signal. All the ANs receive the response signal and form the response-TOA measurement denoted by $\rho_i$. We express it as
\begin{align} \label{eq:rhoANi}
&\rho_i = \nonumber\\
&\left\{
\begin{matrix}
\frac{\left\Vert\boldsymbol{p}_1-\boldsymbol{p}_u\left(t_{TX}^{(u)}\right)\right\Vert}{c} -b_u\left(t_{TX}^{(u)}\right)+ \varepsilon_{1} \text{, } i=1\\
\frac{\left\Vert\boldsymbol{p}_i-\boldsymbol{p}_u\left(t_{TX}^{(u)}\right)\right\Vert}{c}+ b_i\left(t_{TX}^{(u)}\right) -b_u\left(t_{TX}^{(u)}\right)+ \varepsilon_{i} \text{, } i=2,\cdots,M \text{,}
\end{matrix}
\right.
\end{align}
where $t_{TX}^{(u)}$ is the UD transmission time of the response signal.

We observe from (\ref{eq:tauUD}) and (\ref{eq:rhoANi}) that the problem of localization and synchronization for the UD is to estimate the position $\boldsymbol{p}_u$ and the clock offset $b_u$ at a specific instant such as $t_{TX}^{(u)}$ using the measurements $\tau_u$ and $\rho_i$. However, the clock offsets of the SANs $b_i$ are also unknown and need to be estimated as accurate as possible to ensure the accuracy of the UD localization and synchronization. The task of estimating $b_i$ is achieved utilizing the sync-TOA measurements $\tau_i$ in (\ref{eq:tauANiAwo1}) as proposed in Section \ref{virtualsync} and the estimation of $\boldsymbol{p}_u$ and $b_u$ is proposed in Section \ref{locmethod}.

\section{Anchor Virtual Synchronization for PARN}
\label{virtualsync}
In the PARN, we estimate the clock parameters of ANs for UD localization and synchronization without adjusting the AN's clock physically. Therefore, it is referred to as virtual synchronization. The aim of virtual synchronization is to reduce the estimation error of the ANs' clock parameters such that they can contribute to UD localization and synchronization results with high accuracy.

\subsection{Kalman Filter-based Method for Virtual Synchronization}
We note that in the PARN, all the ANs are placed at known positions, and thus the temporal correlation of the clock parameters can be observed from the time series of the sync-TOA measurements received by a SAN. Therefore, in order to track the time-varying clock offset and drift of a SAN, and save computation for consumer-level electronics devices, we present a Kalman filter-based method for each single SAN.

The state vector denoted by $\boldsymbol{x}$ includes clock offset and drift, i.e., $\boldsymbol{x}=[b_i,\omega_i]^T$.
The discrete one-step transition matrix is $\bm{\Phi}$ as given by (\ref{eq:clockbomega}). The measurement matrix is
\begin{align}
\bm{H}=[1 \; 0] \text{.}		
\end{align}

The process noise covariance matrix is $\bm{Q}$ as given by (\ref{eq:Qclock}). The measurement for the filter, denoted by $z$, is
\begin{align}
z(n)=\tau_i(n)-\frac{d_{i1}}{c} \text{,}	
\end{align}
where $n$ is the discrete time index, and the measurement noise variance is $\sigma_i^2$.

The estimation and prediction error variance matrices are denoted by $\bm{P}_{n}$ and $\bm{P}_{n|n-1}$, respectively. The one-step state prediction and estimation are denoted by $\check{\boldsymbol{x}}_{n|n-1}$ and $\check{\boldsymbol{x}}_{n}$, respectively. The entire process of the Kalman filter is given in Appendix \ref{KFprocess}.  The reader can directly use these equations in their applications without looking into related literature.

We note from (\ref{eq:tauANiAwo1}) and (\ref{eq:rhoANi}) that, the instant $t_{TX}^{(u)}$ is not coincident with the TOA measurement time $t_{RX(i)}^{(1)}$ at the SAN. The estimated SAN clock offset from the Kalman filter must be adjusted to the instant $t_{TX}^{(u)}$ in order to be used in UD localization and synchronization. Therefore, we first calculate the time difference to obtain $\Delta t$, i.e., $\Delta t=t_{TX}^{(u)}-t_{RX(i)}^{(1)}$. In practice, the reference time $t$ is not available, and we can use the SAN local reception times to replace $t_{TX}^{(u)}$ and $t_{RX(i)}^{(1)}$. Then we use the state prediction equation (\ref{eq:statepred}) to predict the clock parameters $\check{\boldsymbol{x}}_{n|n-1}$ and the prediction error variance equation (\ref{eq:errorpred}) to predict the error $\bm{P}_{n|n-1}$.

The entire filtering process only involves addition and multiplication of at most $2\times 2$ matrices. The only division operation is applied on a scalar as shown in (\ref{eq:Kalmangain}). This low complexity can easily be achieved by a consumer-level embedded system such as an IoT device. Therefore, this virtual synchronization method is suitable for low-cost electronics systems.

\subsection{Initialization of Kalman Filter}
When initializing the Kalman filter, we need to set the measurement noise variance $\sigma_i^2$, the process noise covariance $\bm{Q}$, the initial state $\check b_i(1)$ and $\check\omega_i(1)$, and the initial error covariance $\bm{P}_1$. The measurement noise $\sigma_i$ is determined by the statistical property of the reception timestamps. The process noise covariance $\bm{Q}$ is related to the quality of the clock source and can be calculated from its Allan variance according to \cite{brown2012introduction}.

Based on (\ref{eq:tauANiAwo1}), we estimate the initial clock offset by subtracting the known propagation time from the sync-TOA measurement, i.e.,
\begin{align} \label{eq:initialb}
\check{b}_{i}(1)=\tau_i(1)-\frac{d_{i1}}{c} \text{.}
\end{align}

By this means, the initial clock offset error is within the measurement noise, and thus, we set the related initial error variance to $\left[\bm{P}_{1}\right]_{1,1}=\sigma_i^2$.

The initial clock drift is computed using the first two sync-TOA measurements, i.e.,
\begin{align} \label{eq:initialome}
\check{\omega}_{i}(1)=\frac{\tau_i(2)-\tau_i(1)}{t_{RX(i)}^{(1)}(2)-t_{RX(i)}^{(1)}(1)}\text{.}
\end{align}

We thereby have the initial error variance for the clock drift as $\left[\bm{P}_{1}\right]_{2,2}=2\sigma_i^2/\left(t_{RX(i)}^{(1)}(2)-t_{RX(i)}^{(1)}(1)\right)^2$. The off-diagonal entries of $\bm{P}_{1}$ are all set to zero.


\section{Optimal UD Localization and Synchronization for PARN} \label{locmethod}
\subsection{TOA Measurement Pre-processing} \label{TOApreprocess}
Employing the virtual synchronization results, we first replace all the clock offsets of the SANs in (\ref{eq:rhoANi}) by their estimations $\check b_i$, omit the time instant $t_{TX}^{(u)}$ for all terms, and then come to
\begin{align} \label{eq:rhoANiE}
\rho_i =\left\{
\begin{matrix}
\frac{\left\Vert\boldsymbol{p}_1-\boldsymbol{p}_u\right\Vert}{c} -b_u+ \varepsilon_{1} \text{, }& i=1\\
\frac{\left\Vert\boldsymbol{p}_i-\boldsymbol{p}_u\right\Vert}{c}+ \check{b}_i -b_u+ \varepsilon_{i}+\varepsilon_{b_i}\text{,} &i=2,\cdots,M \text{.}
\end{matrix}
\right.
\end{align}

By applying the filtering method as presented by Section \ref{virtualsync} on the sync-TOAs given by (\ref{eq:tauANiAwo1}), the estimation of $b_i$ in (\ref{eq:rhoANi}), i.e., $\check{b}_i$, is obtained. The estimation error $\varepsilon_{b_i}$ has a variance determined by the error variance matrix in (\ref{eq:errorpred}), i.e., $\sigma_{b_i}^2=\left[\bm{P}_{n|n-1}\right]_{1,1}$. With a properly designed filter, this error can be significantly reduced compared with the original TOA measurements, as will be shown in Sections \ref{simKalman} and \ref{testKalman}.


The sync-TOA and response-TOA measurements, given by (\ref{eq:tauUD}) and (\ref{eq:rhoANi}), are at different instants. They need to be adjusted to a unified instant. The response delay interval between the UD reception and transmission, denoted by $\delta t$, is $\delta t=t_{TX}^{(u)}-t_{RX(u)}^{(1)}$. Based on the clock model (\ref{eq:clockbomega}) and the UD movement model (\ref{eq:posvel}), we then adjust the time of (\ref{eq:tauUD}) to the UD transmission instant $t_{TX}^{(u)}$, and omit the time instant $t_{TX}^{(u)}$ as
\begin{align} \label{eq:tauUDtxE}
\tau_u = \frac{\left\Vert\boldsymbol{p}_1-\boldsymbol{p}_u+\boldsymbol{v}_u\cdot \delta t\right\Vert}{c}+ b_u - \omega_u \cdot \delta t +\varepsilon_{u} \text{.}
\end{align}

\subsection{Optimal ML Localization and Synchronization}
With these pre-processed measurements given by (\ref{eq:rhoANiE}) and (\ref{eq:tauUDtxE}), we will develop a ML method, namely ML-LAS, to achieve localization and synchronization for a moving UD, in this subsection.

The parameters to be estimated include the UD position $\boldsymbol{p_u}$ and its clock offset $cb_u$ at the UD transmission time $t_{TX}^{(u)}$. The parameters to be estimated are denoted by $\boldsymbol{\theta}$ as
$$
\boldsymbol{\theta}=\left[\boldsymbol{p_u}^T,cb_u\right]^T \text{,}
$$
where the signal propagation speed constant $c$ is included in the clock offset term to unify all the units to meter.

By observing (\ref{eq:tauUDtxE}), we note that the sync-TOA $\tau_u$ is related to the UD velocity $\boldsymbol{v}$ and clock drift $\omega_u$. In other words, we need to handle the UD velocity and clock drift if we use the sync-TOA measurement $\tau_u$ for a moving UD. One idea is that we treat them as known when we use $\tau_u$. Therefore, we will investigate two cases, one is with known velocity and clock drift and another is without. Correspondingly, we develop two modes of the ML-LAS method to deal with the two cases.

For Mode 1, we utilize the sync-TOA $\tau_u$ with known velocity and clock drift. In practice, when the UD is stationary or equipped with motion sensors such as an inertial measurement unit, the UD velocity can be obtained. The UD clock drift can be also estimated when the UD is stationary, as will be shown in Section \ref{experiment}. Therefore, it is both reasonable and practical to treat the UD velocity $\boldsymbol{v}_u$ and clock drift $\omega_u$ as known for Mode 1. For Mode 2, we use the response-TOA measurements only, without $\tau_u$. These measurements are from the response signal transmitted at the same instant and are not related to UD movement or clock drift as shown in (\ref{eq:rhoANiE}). Therefore, Mode 2 is also suitable for a moving UD.


For the two modes, the measurements are written in the collective form as
$$
\boldsymbol{\rho}=\left\{
\begin{matrix}
\left[c\rho_1,\cdots,c\rho_M,c\tau_u\right]^T \text{,} &\text{for Mode 1}\\
\left[c\rho_1,\cdots,c\rho_M\right]^T \text{,} &\text{for Mode 2}
\end{matrix}
\right.
$$

The relation between the unknown parameters and the measurements is
\begin{equation} \label{eq:rhoandtheta}
\boldsymbol{\rho} = h(\boldsymbol{\theta}) + \boldsymbol{\varepsilon} \text{,}
\end{equation}
where according to (\ref{eq:rhoANiE}) and (\ref{eq:tauUDtxE}), the $i$-th row of the function $h(\boldsymbol{\theta})$ is
\begin{align} \label{eq:funtheta}
&\left[h(\boldsymbol{\theta})\right]_{i,:} = \nonumber\\
&\left\{
\begin{matrix}
\left\Vert\boldsymbol{p}_1-\boldsymbol{p}_u\right\Vert-cb_u, & i=1 \\
\left\Vert\boldsymbol{p}_i-\boldsymbol{p}_u\right\Vert+c\check b_i-cb_u, & i=2,\cdots,M \\
\left\Vert\boldsymbol{p}_1-\boldsymbol{p}_u+\boldsymbol{v}_u\cdot\delta t\right\Vert+cb_u-c\omega_u \delta t,&i=M+1
\text{,}
\end{matrix} 
\right.
\end{align}
with $\boldsymbol{\varepsilon}=\left[c\varepsilon_1,c\varepsilon_{b_2}+c\varepsilon_2,\cdots,c\varepsilon_{b_M}+c\varepsilon_M,c\varepsilon_u\right]^T$, for Mode 1,
and
\begin{align} \label{eq:funthetacase2}
\left[h(\boldsymbol{\theta})\right]_{i,:} = 
\left\{
\begin{matrix}
\left\Vert\boldsymbol{p}_1-\boldsymbol{p}_u\right\Vert-cb_u, & i=1 \\
\left\Vert\boldsymbol{p}_i-\boldsymbol{p}_u\right\Vert+c\check b_i-cb_u, & i=2,\cdots,M
\text{,}
\end{matrix} 
\right.
\end{align}
with $\boldsymbol{\varepsilon}=\left[c\varepsilon_1,c\varepsilon_{b_2}+c\varepsilon_2,\cdots,c\varepsilon_{b_M}+c\varepsilon_M\right]^T$, for Mode 2.

Recall that all the error terms in the measurements are independently Gaussian distributed. The ML estimation of $\boldsymbol{\theta}$ is equivalent to solving the weighted least squares (WLS) minimization problem as
\begin{equation} \label{eq:MLminimizer}
\hat{\boldsymbol{\theta}}=\text{arg}\min\limits_{{\boldsymbol{\theta}}} \left\Vert\boldsymbol{\rho} - \mathit{h}({\boldsymbol{\theta}})\right\Vert_{\bm{W}}^2
\text{,}
\end{equation}
where $\hat{\boldsymbol{\theta}}$ is the estimator, and $\bm{W}$ is a diagonal positive-definite weighting matrix given by
\begin{equation} \label{eq:matW}
\bm{W}=\left\{
\begin{matrix}
\left[
	\begin{matrix}
	\bm{W}_0 & \boldsymbol{0}_{M}\\
	\boldsymbol{0}_{M}^T& \frac{1}{c^2\sigma_u^2}
	\end{matrix}
\right] \text{,} & \text{for Mode 1}\\
\bm{W}_0 \text{,} &\text{for Mode 2}
\end{matrix}
\right.
\end{equation}
in which $\boldsymbol{0}_{M}$ is a vector with all zero elements, and
\begin{equation} \label{eq:matW0}
\bm{W}_0=\mathrm{diag}\left(\frac{1}{c^2\sigma_1^2},\frac{1}{c^2\sigma_{b_2}^2+c^2\sigma_2^2},\cdots,\frac{1}{c^2\sigma_{b_M}^2+c^2\sigma_M^2}\right) \text{.}
\end{equation}

\textbf{Remark 1}: Note from (\ref{eq:matW0}) that the estimation error is related to the error term $\sigma_{b_i}$. It is determined by the virtual synchronization for SAN. Utilizing historical measurements by a Kalman filter can reduce the value of $\sigma_{b_i}$, improving the estimation accuracy for $\boldsymbol{\theta}$.

\subsection{Gauss-Newton Localization and Synchronization Algorithm}
Following \cite{kaplan2005understanding,misra2006global}, we apply the Gauss-Newton method to construct the iterative localization algorithm to solve the minimization problem given by (\ref{eq:MLminimizer}).

We convert the non-linear relation between the measurements and the parameters to a linear one by conducting a Taylor series expansion on (\ref{eq:rhoandtheta}) at the estimate point of
$
\check{\boldsymbol{\theta}}=\left[\check{\boldsymbol{p}}_u^T,c\check{b}_u\right]^T \text{,}
$
where $\check{\boldsymbol{p}}_u$ and $\check{b}_u$ are estimates for $\boldsymbol{p}_u$ and $b_u$, respectively.

If the parameter estimation is close to the true value, the first-order term is retained and the higher order terms are ignored. Then, (\ref{eq:rhoandtheta}) becomes
\begin{equation} \label{eq:GNtaylor}
	\boldsymbol{\rho} = \mathit{h}(\check{\boldsymbol{\theta}}) + \left(\frac{\partial \mathit{h}(\boldsymbol{\theta})}{\partial \boldsymbol{\theta}}|_{\boldsymbol{\theta}=\check{\boldsymbol{\theta}}}\right)\left(\boldsymbol{\theta}-\check{\boldsymbol{\theta}}\right)+\boldsymbol\varepsilon \text{.}
\end{equation}

We denote the error vector by $\Delta\boldsymbol{\theta}$, and 
$\Delta\boldsymbol{\theta} = \boldsymbol{\theta}-\check{\boldsymbol{\theta}} \text{.}
$

The design matrix is denoted by $\bm{G}$,
\begin{align} \label{eq:matG}
\bm{G}=\frac{\partial \mathit{h}(\boldsymbol{\theta})}{\partial \boldsymbol{\theta}}|_{\boldsymbol{\theta}=\check{\boldsymbol{\theta}}} =\left\{
\begin{matrix}
\left[
\begin{matrix}
\bm{G}_0\\
\boldsymbol{g}^T
\end{matrix}
\right] \text{,} &\text{for Mode 1}\\
\bm{G}_0 \text{,} &\text{for Mode 2}
\end{matrix}
\right.
\end{align}
where
\begin{equation} \label{eq:matG0}
\bm{G}_0=\left[
\begin{matrix}
-\boldsymbol{e}_1^T & -1 \\
\vdots & \vdots \\
-\boldsymbol{e}_M^T & -1
\end{matrix}
\right]\text{,}
\end{equation}
with $\boldsymbol{e}$ representing the unit line-of-sight (LOS) vector from the UD to the AN at the time instant of UD transmission,
\begin{equation} \label{eq:rhoLOS}
\boldsymbol{e}_{i}=\frac{\boldsymbol{p}_i - \check{\boldsymbol{p}}_u }{\Vert \boldsymbol{p}_i - \check{\boldsymbol{p}}_u\Vert } , i=1,\cdots,M,
\end{equation}
and
$$
\boldsymbol{g}=\left[
\begin{matrix}
-\boldsymbol{l}^T & 1
\end{matrix}
\right]^T\text{,}
$$
with $\boldsymbol{l}$ representing the unit LOS vector from the UD to AN \#1 at the UD reception time as
\begin{equation} \label{eq:tauLOS}
\boldsymbol{l}=\frac{\boldsymbol{p}_1 - \check{\boldsymbol{p}}_u+\boldsymbol{v}_u \delta t}{\Vert \boldsymbol{p}_1 - \check{\boldsymbol{p}}_u+\boldsymbol{v}_u \delta t\Vert } \text{.}
\end{equation}

The residual vector is denoted by $\boldsymbol{r}$, 
\begin{equation}\label{eq:residual}
\boldsymbol{r} = \boldsymbol{\rho} - \mathit{h}(\check{\boldsymbol{\theta}})= \bm{G} \cdot \Delta\boldsymbol{\theta}+\boldsymbol\varepsilon \text{.}
\end{equation}

The WLS estimate of the error vector $\Delta\boldsymbol{\theta}$ is denoted by $\Delta \check{\boldsymbol{\theta}}$,
\begin{equation} \label{eq:leastsquare}
\Delta\check{\boldsymbol{\theta}}=(\bm{G}^T\bm{W}\bm{G})^{-1}\bm{G}^T\bm{W}\boldsymbol{r} \text{.}
\end{equation}

The estimated parameter vector is thereby updated iteratively by
\begin{equation} \label{eq:esttheta}
\check{\boldsymbol{\theta}} \leftarrow \check{\boldsymbol{\theta}} + \Delta \check{\boldsymbol{\theta}} \text{.}
\end{equation}

The design matrix $\bm{G}$ and the residual $\boldsymbol{r}$ are updated by the estimated parameter from (\ref{eq:esttheta}) iteratively until convergence.

\section{Localization and Synchronization Performance Analysis} \label{locanalysis}

\subsection{Localization and Synchronization Error} \label{ErrorAnalysis}


We denote the bias of the parameter estimation $\check{\boldsymbol{\theta}}$ by $\boldsymbol{\mu}$. According to \cite{kay1993fundamentals}, a ML estimator is asymptotically unbiased. Therefore,
\begin{equation} \label{eq:bias}
\boldsymbol{\mu}=\mathbb{E}[\Delta\boldsymbol{\theta}]=\boldsymbol{0} \text{.}
\end{equation}

The estimation error variance denoted by $\bm{\Lambda}$ is given by
\begin{align} \label{eq:matQ}
\bm{\Lambda}=\mathbb{E}\left[\left(\Delta\boldsymbol{\theta}-\mathbb{E}[\Delta\boldsymbol{\theta}]\right)\left(\Delta\boldsymbol{\theta}-\mathbb{E}[\Delta\boldsymbol{\theta}]\right)^T\right] =(\bm{G}^T\bm{W}\bm{G})^{-1}\text{,} 
\end{align}
and the root mean square error ($RMSE$) is
\begin{align}
RMSE =\sqrt{\Vert\boldsymbol{\mu}\Vert^2+\mathrm{tr}(\bm{\Lambda})}=\sqrt{\mathrm{tr}(\bm{\Lambda})}.
\end{align}

\subsection{CRLB Derivation}\label{CRLBanalysis}
CRLB is a commonly used metric to evaluate the error of an unbiased estimator. We derive the CRLB of the proposed ML-LAS method in this subsection.

With the TOA measurements collected in the PARN, the likelihood function, denoted by $f(\boldsymbol{\rho}|\boldsymbol{\theta})$, is
\begin{equation} \label{eq:likelihood}
f(\boldsymbol{\rho}|\boldsymbol{\theta}) = \frac{\exp\left(-\frac{1}{2} \Vert\boldsymbol{\rho} - \mathit{h}({\boldsymbol{\theta}})\Vert_{\bm{W}}^2\right)}{(2\pi)^{\frac{\Gamma}{2}} |\bm{W}^{-1}|^{\frac{1}{2}}} \text{,}
\end{equation}
where $\Gamma=M+1$ for Mode 1 and $\Gamma=M$ for Mode 2.


We denote the Fisher information matrix (FIM) by $\mathcal{F}$, 
\begin{equation} \label{eq:FisherExpectation}
\mathcal{F}=-\mathbb{E}\left[\frac{\partial^2 \ln f(\boldsymbol{\rho}|\boldsymbol{\theta})}{\partial\boldsymbol{\theta}^T \partial\boldsymbol{\theta}}\right]=\left(\frac{\partial h(\boldsymbol{\theta})}{\partial \boldsymbol{\theta}}\right)^T\bm{W}\frac{\partial h(\boldsymbol{\theta})}{\partial \boldsymbol{\theta}} \text{.}
\end{equation}

Based on (\ref{eq:matG}), the FIM is thereby
\begin{equation}\label{eq:FIMvsG}
\mathcal{F} = \bm{G}^T\bm{W}\bm{G} \text{.}
\end{equation}

The CRLB relating to the $i$-th element in the parameter vector $\boldsymbol{\theta}$  is then obtained by 
\begin{equation} \label{eq:CRLBFisher}
\mathsf{CRLB}([\boldsymbol{\theta}]_i)=[\mathsf{CRLB}]_{i,i}=[\mathcal{F}^{-1}]_{i,i} \text{,}
\end{equation}
where  $[\cdot]_{i}$ represents the $i$-th element of a vector, and $[\cdot]_{i,j}$ represents the entry at the $i$-th row and the $j$-th column of a matrix.


We investigate the CRLB in the two modes, i.e., Mode 1 (with the sync-TOA $\tau_u$) and Mode 2 (without). We denote the FIMs of the two modes by $\mathcal{F}_1$ for Mode 1 and $\mathcal{F}_2$ for Mode 2, respectively.

Based on (\ref{eq:matG}) and (\ref{eq:CRLBFisher}), we have
\begin{align} \label{eq:FIMandFIM0}
\mathcal{F}_1& = 
\left[\begin{matrix}
\bm{G}_0^T & \boldsymbol{g}
\end{matrix}\right]
\left[\begin{matrix}
\bm{W}_0 & \boldsymbol{0}_M\\
\boldsymbol{0}_M & \frac{1}{c^2\sigma_u^2}
\end{matrix}\right]
\left[\begin{matrix}
\bm{G}_0 \\
\boldsymbol{g}^T
\end{matrix}\right] \nonumber\\
&= \bm{G}_0^T\bm{W}_0\bm{G}_0+\frac{1}{c^2\sigma_u^2}\boldsymbol{g}\boldsymbol{g}^T 
\text{.}
\end{align}


We note that 
\begin{equation}\label{eq:matFIM0}
\mathcal{F}_2=\bm{G}_0^T\bm{W}_0\bm{G}_0 \text{.}
\end{equation}
Therefore, by substituting (\ref{eq:matFIM0}) into (\ref{eq:FIMandFIM0}) we come to
\begin{align} \label{eq:FIM1andFIM2}
\mathcal{F}_1
&=\mathcal{F}_2+\frac{1}{c^2\sigma_u^2}\boldsymbol{g}\boldsymbol{g}^T
\text{.}
\end{align}


We can see from (\ref{eq:FIM1andFIM2}) that 
\begin{equation} \label{eq:compareFIM}
\mathcal{F}_1\succeq\mathcal{F}_2 \text{,}
\end{equation}
which, according to (\ref{eq:CRLBFisher}), shows that the CRLB of Mode 1, denoted by $\mathsf{CRLB}_1$ is smaller than the CRLB of Mode 2, denoted by $\mathsf{CRLB}_2$, i.e.,
\begin{equation} \label{eq:compareCRLB}
[\mathsf{CRLB}_1]_{i,i}\leq [\mathsf{CRLB}_2]_{i,i} \text{.}
\end{equation}

\textbf{Remark 2}: With known UD velocity and clock drift, utilizing the sync-TOA $\tau_u$ in Mode 1 brings higher localization and synchronization accuracy than Mode 2.

The two modes of the proposed ML-LAS method can be applied to different cases. For example, if the velocity and clock drift of the UD can be obtained, the ML-LAS method can work in Mode 1 to achieve best estimation accuracy. For cases in which the UD does not transfer the sync reception time information in the response signal to save power, or the UD velocity and/or clock drift is not easy to obtain due to very low-cost oscillator or very high dynamic maneuver, Mode 2 is more suitable.


\subsection{Impact of UD Velocity and Clock Drift Deviation on ML-LAS Mode 1} \label{deviatedV}
\subsubsection{Error caused by Velocity Deviation}
In a real system, the UD velocity information may not be accurate. We investigate how the deviation of the velocity impacts the final localization and synchronization results for Mode 1 of the ML-LAS method. With this as a guidance, we can determine in what cases the proposed ML-LAS method can work in Mode 1 and predict how much error will be produced.

The known UD velocity we obtained in such a case is denoted by $\tilde{\boldsymbol{v}}_u$. The deviation from the true velocity is denoted by $\Delta \boldsymbol{v}_u=\tilde{\boldsymbol{v}}_u-\boldsymbol{v}_u$. The deviated velocity-caused measurement error vector is denoted by ${\boldsymbol{r}}_v$ and is given by
\begin{align}\label{eq:resdV}
{\boldsymbol{r}}_v 
=\left[
\begin{matrix}
\boldsymbol{0}_M\\
\left\Vert \boldsymbol{p}_1 - \boldsymbol{p}_u + \tilde{\boldsymbol{v}}_u  \delta t\right\Vert  -\left\Vert \boldsymbol{p}_1 - {\boldsymbol{p}}_u + {\boldsymbol{v}}_u  \delta t\right\Vert
\end{matrix}
\right]
\text{,}
\end{align}

Then, the estimation bias, denoted by ${\boldsymbol{\mu}}_v$, is 
\begin{equation}\label{eq:dPvsdV}
{\boldsymbol{\mu}}_v=({\bm{G}}_v^T\bm{W}{\bm{G}}_v)^{-1}{\bm{G}}_v^T\bm{W}{\boldsymbol{r}}_v \text{,}
\end{equation}
where the design matrix ${\bm{G}}_v$ is
\begin{equation} \label{eq:GdV}
[{\bm{G}}_v]_{i,:}=
\left\{
\begin{array}{ll}
\left[-\boldsymbol{e}_i^T , -1\right]\text{,}\; &i=1,\cdots,M\\
\left[-{\boldsymbol{l}}_v^T ,1\right]\text{,}\; &i=M+1\text{,}
\end{array}
\right.
\end{equation}
with the LOS vector ${\boldsymbol{l}}_v$ that reads
\begin{equation} \label{eq:tauLOSdV}
{\boldsymbol{l}}_v=\frac{\boldsymbol{p}_1 - {\boldsymbol{p}_u}+\tilde{\boldsymbol{v}}_u \delta t}{\Vert \boldsymbol{p}_1 - {\boldsymbol{p}_u}+\tilde{\boldsymbol{v}}_u \delta t\Vert } \text{.}
\end{equation}

We then come to
\begin{equation} \label{eq:mudvsquare}
	\Vert{\boldsymbol{\mu}}_v\Vert^2={\boldsymbol{r}}_v^T\bm{S}_v^T\bm{S}_v{\boldsymbol{r}}_v \text{,}
\end{equation}
where $\bm{S}_v=({\bm{G}}_v^T\bm{W}{\bm{G}}_v)^{-1}{\bm{G}}_v^T\bm{W}$.

By observing (\ref{eq:resdV}), we note that only the last element of ${\boldsymbol{r}}_v$ is non-zero. Therefore, (\ref{eq:mudvsquare}) becomes
\begin{equation} \label{eq:mudvsquare1}
	\Vert{\boldsymbol{\mu}}_v\Vert^2=\left[\bm{S}_v^T\bm{S}_v\right]_{M+1,M+1}\left[{\boldsymbol{r}}_v\right]_{M+1}^2 \text{.}
\end{equation}

We also note that if the measurement error is small and the UD is far from ANs, we have
$$
\left\Vert \boldsymbol{p}_1 - \boldsymbol{p}_u + \tilde{\boldsymbol{v}}_u  \delta t\right\Vert  -\left\Vert \boldsymbol{p}_1 - {\boldsymbol{p}}_u + {\boldsymbol{v}}_u  \delta t\right\Vert={\boldsymbol{l}}_v^T\Delta\boldsymbol{v}_u\delta t 
\text{.}
$$
Therefore, 
\begin{equation} \label{eq:mudvsquare2}
	\Vert{\boldsymbol{\mu}}_v\Vert^2=\left[\bm{S}_v^T\bm{S}_v\right]_{M+1,M+1} {\boldsymbol{l}}_v^T{\boldsymbol{l}}_v \Vert\Delta\boldsymbol{v}_u\Vert^2\delta t^2
	\text{.}
\end{equation}

We observe from (\ref{eq:mudvsquare2}) that the estimation bias increases linearly with the growth of the product of the norm of UD velocity deviation and the response delay time, i.e., $\delta t\Vert\Delta \boldsymbol{v}_u\Vert$.

The estimation error variance caused by the measurement noise, denoted by ${\bm{\Lambda}}_v$, is
\begin{equation} \label{eq:posvardV}
{\bm{\Lambda}}_v=({\bm{G}}_v^T\bm{W}{\bm{G}}_v)^{-1} \text{.}
\end{equation}

The RMSE, denoted by $RMSE_v$, is
\begin{equation} \label{eq:RMSEdV}
	RMSE_v  =\sqrt{\Vert{\boldsymbol{\mu}}_v\Vert^2+\mathrm{tr}({\bm{\Lambda}}_v)} \text{.}
\end{equation}


\subsubsection{Error caused by UD Clock Drift Deviation} \label{deviateOme}
Another case is that the UD clock drift is not accurately known and has deviation from the true value. This will cause an estimation error to Mode 1 of the proposed ML-LAS method. We denote the clock drift that may contain deviation by $\tilde{\boldsymbol{\omega}}_u$. The deviation is denoted by $\Delta \omega_u=\tilde{\boldsymbol{\omega}}_u-\omega_u$. The clock drift-caused measurement error vector, denoted by ${\boldsymbol{r}}_{\omega}$, is
\begin{align}\label{eq:resdome}
{\boldsymbol{r}}_{\omega}
=\left[
\begin{matrix}
\boldsymbol{0}_M\\
c\Delta \omega_u\delta t
\end{matrix}
\right]
\text{,}
\end{align}

Then, the estimation bias, denoted by ${\boldsymbol{\mu}}_\omega$, is 
\begin{equation}\label{eq:dPvsdome}
{\boldsymbol{\mu}}_\omega=({\bm{G}}^T\bm{W}{\bm{G}})^{-1}{\bm{G}}^T\bm{W}{\boldsymbol{r}}_\omega \text{,}
\end{equation}
where the design matrix ${\bm{G}}$ is given by (\ref{eq:matG}), and $\bm{W}$ is given by (\ref{eq:matW}).

Similar to the deviated UD velocity case, we have
\begin{equation} \label{eq:mudomesquare}
\Vert{\boldsymbol{\mu}}_\omega\Vert^2=\left[\bm{S}_\omega^T\bm{S}_\omega\right]_{M+1,M+1}  c^2\Delta\omega_u^2\delta t^2 
\text{.}
\end{equation}
where $\bm{S}_\omega=({\bm{G}}^T\bm{W}{\bm{G}})^{-1}{\bm{G}}^T\bm{W}$.

We show from (\ref{eq:mudomesquare}) that the estimation bias increases linearly with a growing product of the UD clock drift deviation and the response delay time, i.e., $|\Delta \omega_u|\delta t$.

The estimation error variance is also $\bm{\Lambda}$ given by (\ref{eq:matQ}). The RMSE, denoted by $RMSE_\omega$, is
\begin{equation} \label{eq:RMSEdome}
	RMSE_\omega  =\sqrt{\Vert\boldsymbol{\mu}_\omega\Vert^2+\mathrm{tr}({\bm{\Lambda}})} \text{.}
\end{equation}

\section{Numerical Simulation} \label{simulation}
Numerical simulations are conducted in this section to evaluate the virtual synchronization and the localization and synchronization performance of the proposed PARN. In all the simulations, we compute the RMSE of the positioning and timing results. The CRLB is used as a metric to evaluate the estimation accuracy. The RMSE of the positioning results is
\begin{align} \label{eq:RMSEpos}
\text{Position RMSE}&=\sqrt{\frac{1}{N_s}\sum_{1}^{N_s}\Vert\boldsymbol{p}_u-\hat{\boldsymbol{p}}_u\Vert^2} \text{,}
\end{align}
where $N_s$ is the total number of positioning result samples from the simulation, and $\hat{\boldsymbol{p}}_u$ is the localization result from the proposed algorithm for each simulated sample.

The RMSE of the clock offset results is
\begin{align} \label{eq:RMSEclockb}
\text{Clock offset RMSE}&=\sqrt{\frac{1}{N_s}\sum_{1}^{N_s}\left(cb_u-c\hat{b}_u\right)^2} \text{.}
\end{align}

\begin{figure}
	\centering
	\includegraphics[width=0.99\linewidth]{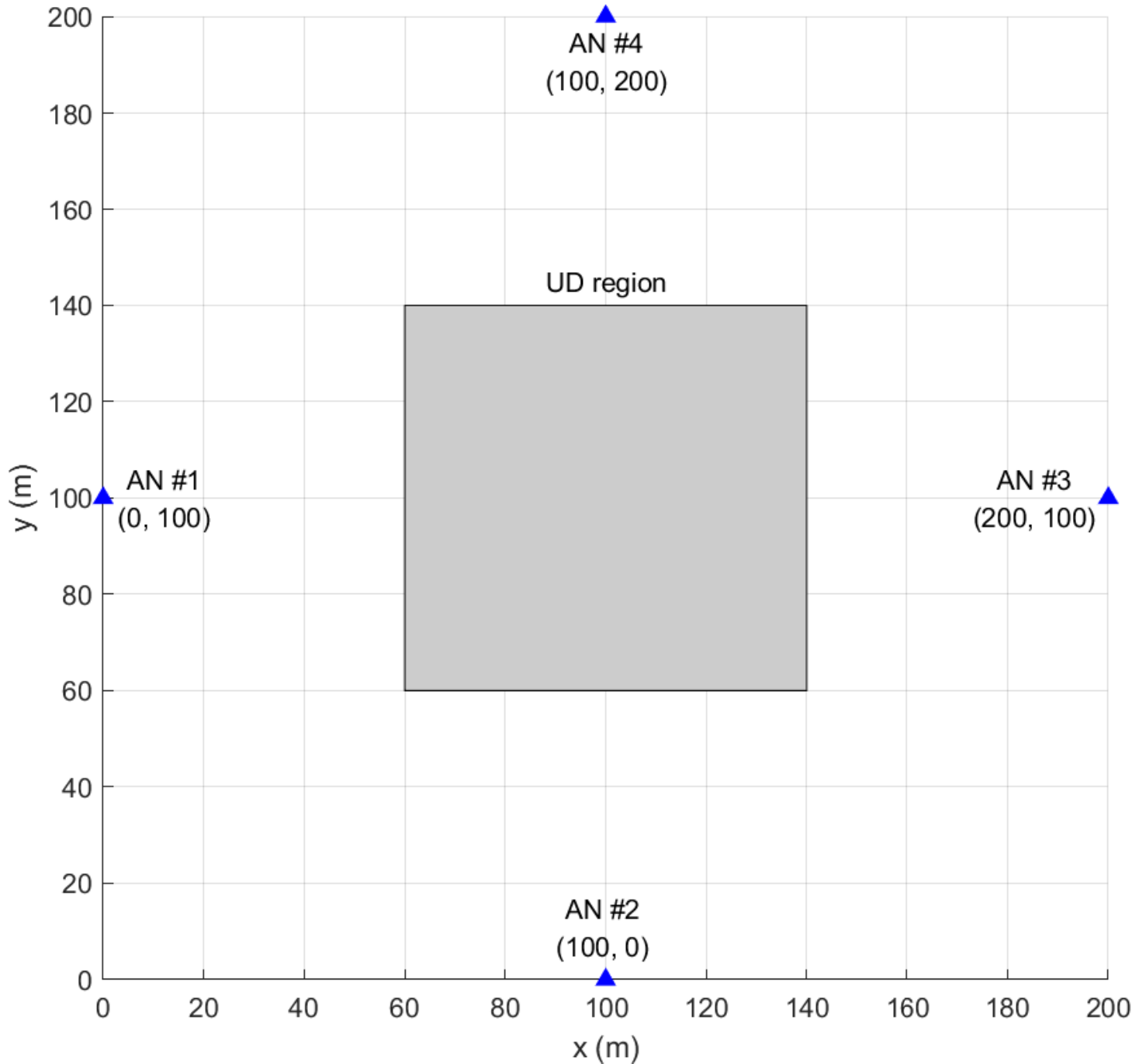}
	\caption{ANs and UD positions for simulation. Four ANs are placed at the sides of the square area and UD is moving randomly in the gray square region. }
	\label{fig:simulationsetting}
\end{figure}

A simulation scene for the PARN is created to evaluate the performance of the proposed virtual synchronization and ML-LAS method in different cases. Four ANs are placed on the middle of the four sides of a 200 m$\times$200 m square area as shown in Fig. \ref{fig:simulationsetting}. AN \#1 is set as the PAN and other ANs are SANs. The total simulation time length is set to 100 s. AN \#1 periodically transmits the sync signal with an interval of 10 ms. The UD receives the sync signal, and after a 5 ms delay transmits the response signal that is received by all ANs.

The UD is placed inside the square area with a side length of 80 m and center at (100, 100) m, as shown in Fig. \ref{fig:simulationsetting}. During each 10 ms interval, the UD moves from a random start point inside this area with a speed of 5 m/s and a random direction. Thus, the number of the 10-ms intervals is 10,000. The clock offset and drift of the UD are set randomly during each different 10-ms time interval. The UD clock offset follows a uniform distribution $b_u\sim \mathcal{U}(-1,1)$ s, and the clock drift follows $\omega_u\sim \mathcal{U}(-20,20)$ ppm, which follow the requirements in \cite{IEEE802.15.4-2020}.

The clock offsets and drifts of the SANs change with time following the clock model (\ref{eq:clockbomega}). We set the value of the clock model parameters following \cite{brown2012introduction}. The signal propagation speed $c$ is set to the speed of light. The main simulation settings are listed in Table \ref{table_simsetting}.

\begin{table}[!t]
		\caption{Simulation settings}
		\label{table_simsetting}
		\centering
		\begin{tabular}{>{\centering\arraybackslash}m{2cm} >{\centering\arraybackslash}m{4cm} >{\centering\arraybackslash}m{1.2cm} }
			\toprule
			Parameter & Value & Unit\\
			
			\hline
			PAN  & AN \#1 & -\\
			SAN  & AN \#2, \#3, \#4 & -\\
			AN/UD position & as given by Fig. \ref{fig:simulationsetting}& -\\
			Initial $b_i$ &\#2: $-5\times 10^{-7}$, \#3: $8\times 10^{-8}$, \#4: $2\times 10^{-1}$ & s\\
			Initial $\omega_i$ &\#2: 1, \#3: 5, \#4:-3& ppm\\
			$\Vert\boldsymbol{v}_u \Vert$ & 5 & m/s\\
			$s_b$ & $10^{-21}$ & s\\
			$s_{\omega}$ & $5.9\times 10 ^{-23}$ & s$^{-1}$\\
			$b_u$ & $\mathcal{U}(-1,1)$ & s\\
			$\omega_u$ & $\mathcal{U}(-20,20)$ & ppm\\
			\bottomrule
		\end{tabular}
		
\end{table}

\subsection{Kalman Filter-based Virtual Synchronization for SANs} \label{simKalman}
We investigate the SAN's clock offset estimation error at the time of the transmission of the UD response signal because the clock offset at this moment is used in UD localization and synchronization. The sync-TOA measurement noise $c\sigma_i$ at the AN end is set to 0.05 m, which is at the level of a UWB ranging system \cite{ruiz2017comparing}. The true clock offset at every instant is simulated based on the clock model given by (\ref{eq:clockbomega}) and the parameters given by Table \ref{table_simsetting}. The sync-TOA measurements $\tau_i$ are generated based on (\ref{eq:tauANiAwo1}). The Kalman filter presented by Section \ref{virtualsync} is applied to estimate the clock offset.

\begin{figure}
	\centering
	\includegraphics[width=0.99\linewidth]{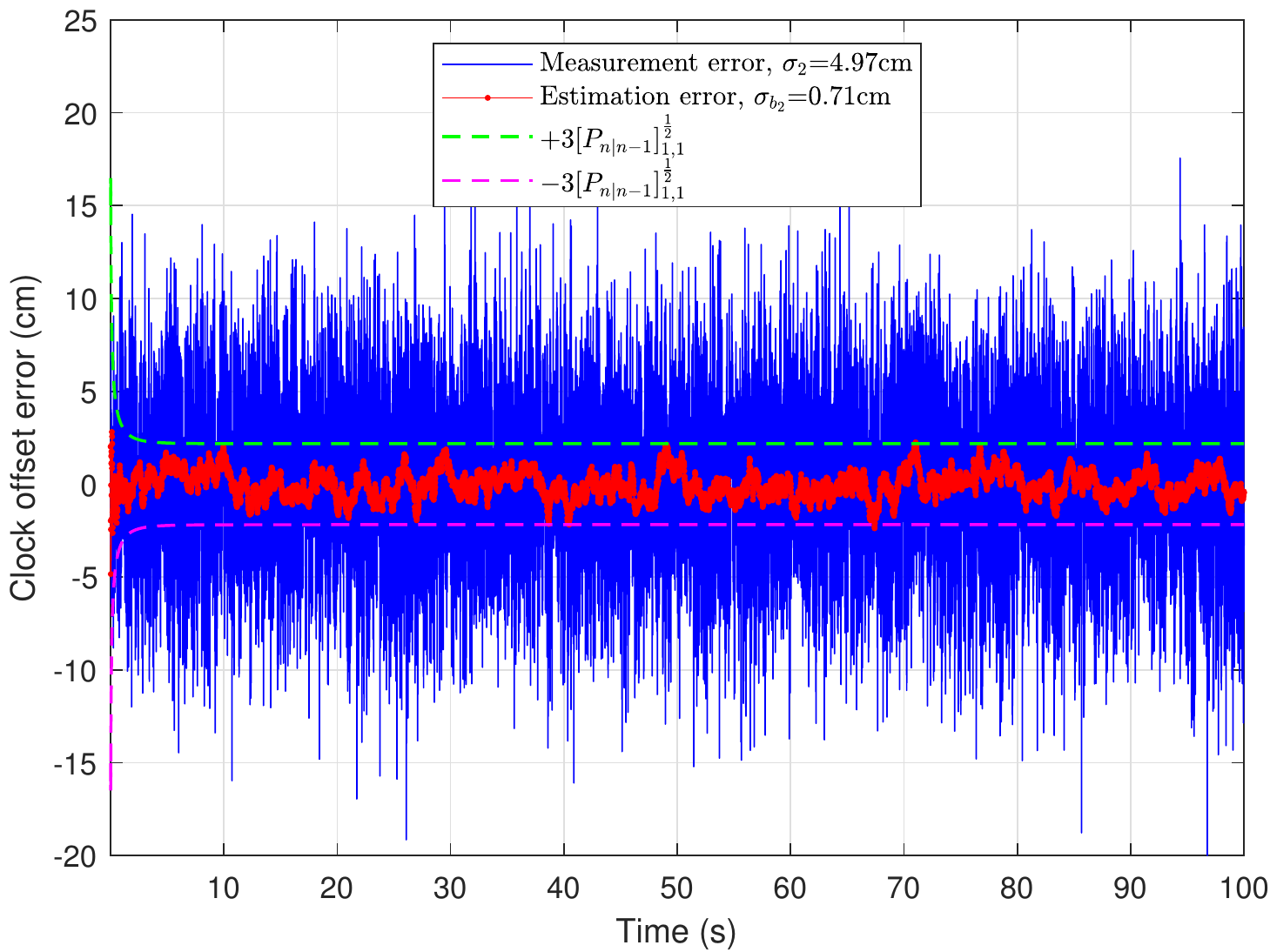}
	\caption{Virtual synchronization results for SANs from numerical simulation. The clock offset error estimated by the proposed Kalman filter is significantly reduced compared with the measurement error, and is within the estimated error variance derived from $\bm{P}_{n|n-1}$, i.e., $\pm3[\bm{P}_{n|n-1}]_{1,1}^{\frac{1}{2}}$, showing the effectiveness of the proposed filter. }
	\label{fig:KFresult}
\end{figure}

We take the clock offset estimation result of AN \#2 shown in Fig. \ref{fig:KFresult} as an example. We can see that after filtering, the clock offset error is significantly smaller than the original measurement error, showing the advantage over the existing methods using one-time measurements only. The $\sigma_{b_2}$ of the estimation error is bounded by the error variance $\bm{P}_{n|n-1}$ computed by (\ref{eq:errorpred}) from the filter. The value of $[\bm{P}_{n|n-1}]_{1,1}^{\frac{1}{2}}$ is 0.73 cm, approximately equal to the estimation error $\sigma_{b_2}$. The results for other ANs are similar. This validates the effectiveness of the proposed Kalman filter-based method in estimating the SAN's clock offset correctly with reduced error, and shows that $[\bm{P}_{n|n-1}]_{1,1}$ can be used as the error variance $\sigma_{b_i}^2$ in the ML-LAS method, consistent with Section \ref{TOApreprocess}.

\subsection{UD Localization and Synchronization Performance}
We set that the TOA measurement noise $c\sigma_u$ and $c\sigma_i$ are identical, and vary from 0.01 m to 1 m with six steps in total. In every step, 10,000 times of Monte-Carlo simulations are done to generate the random UD position and motion, the clock offset and drift, and the TOA measurements. The simulated data are input to the Gauss-Newton ML-LAS algorithm proposed in Section \ref{locmethod}. The maximum iteration time for the localization algorithm is set to 10.

\begin{figure}
	\centering
	\includegraphics[width=0.99\linewidth]{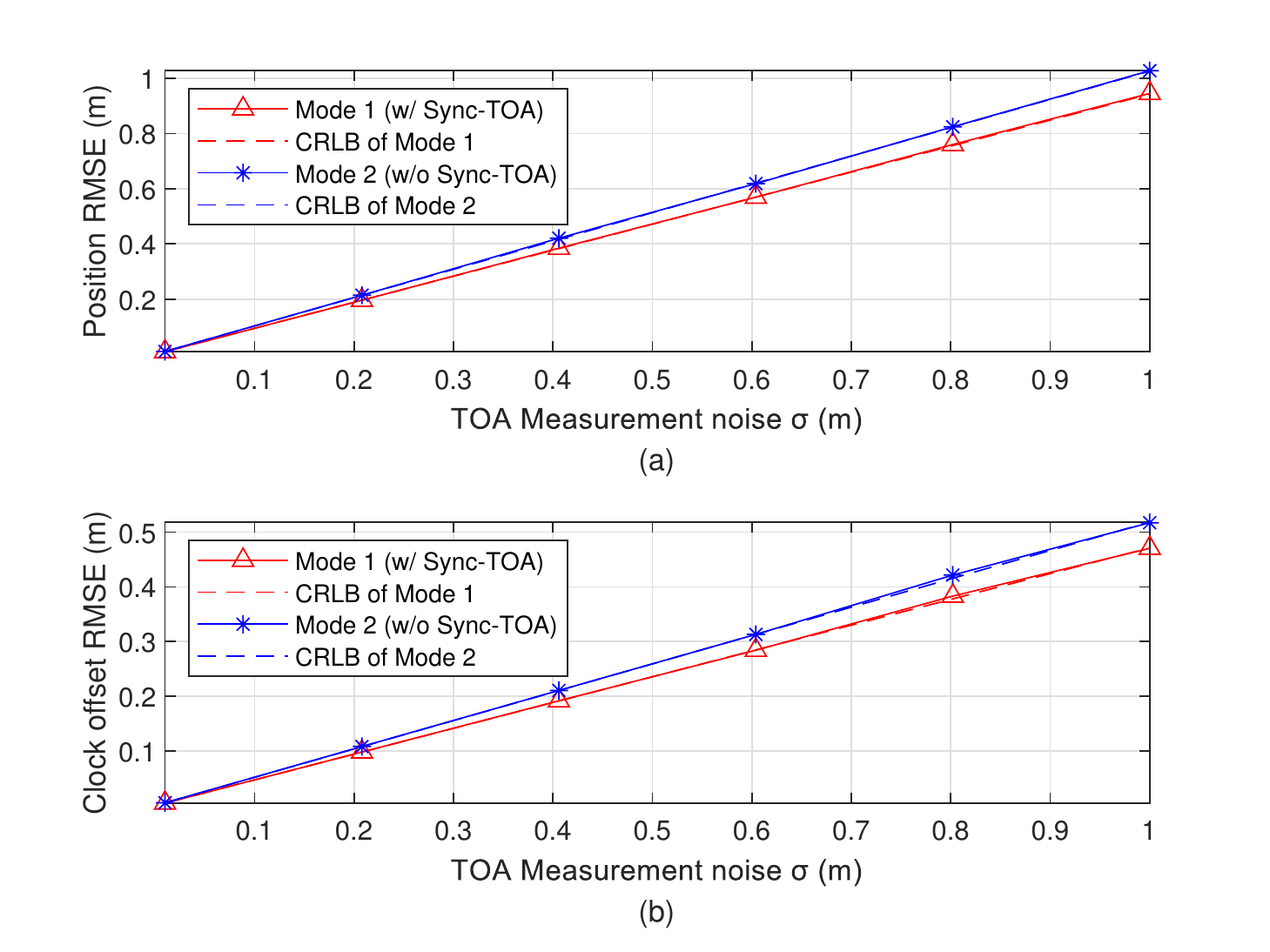} 
	\caption{Localization and synchronization error vs. measurement noise from numerical simulation. The position and clock offset estimation errors of the proposed ML-LAS method reach CRLB. The error of Mode 1 (using sync-TOA) is smaller than that of Mode 2 (without using sync-TOA).
	}
	\label{fig:compareTOA}
\end{figure}

The UD position and clock offset estimation results of the two modes (Mode 1 with sync-TOA and Mode 2 without) are shown in Fig. \ref{fig:compareTOA}. The localization error result of both modes along with their respective CRLBs are shown in Fig. \ref{fig:compareTOA} (a). It can be observed that the theoretical CRLB is identical with the numerical position RMSE result given by the ML-LAS method. The clock offset results are shown in Fig. \ref{fig:compareTOA} (b). The estimation errors also reach their respective CRLBs. This figure indicates that the proposed ML-LAS method is an unbiased estimator. Comparing the two modes of the proposed ML-LAS method, we can see that the CRLB of Mode 1 (with sync-TOA $\tau_u$) is smaller than that of Mode 2 without using $\tau_u$, consistent with the theoretical analysis in Section \ref{CRLBanalysis}.


\begin{figure}
	\centering
	\includegraphics[width=0.99\linewidth]{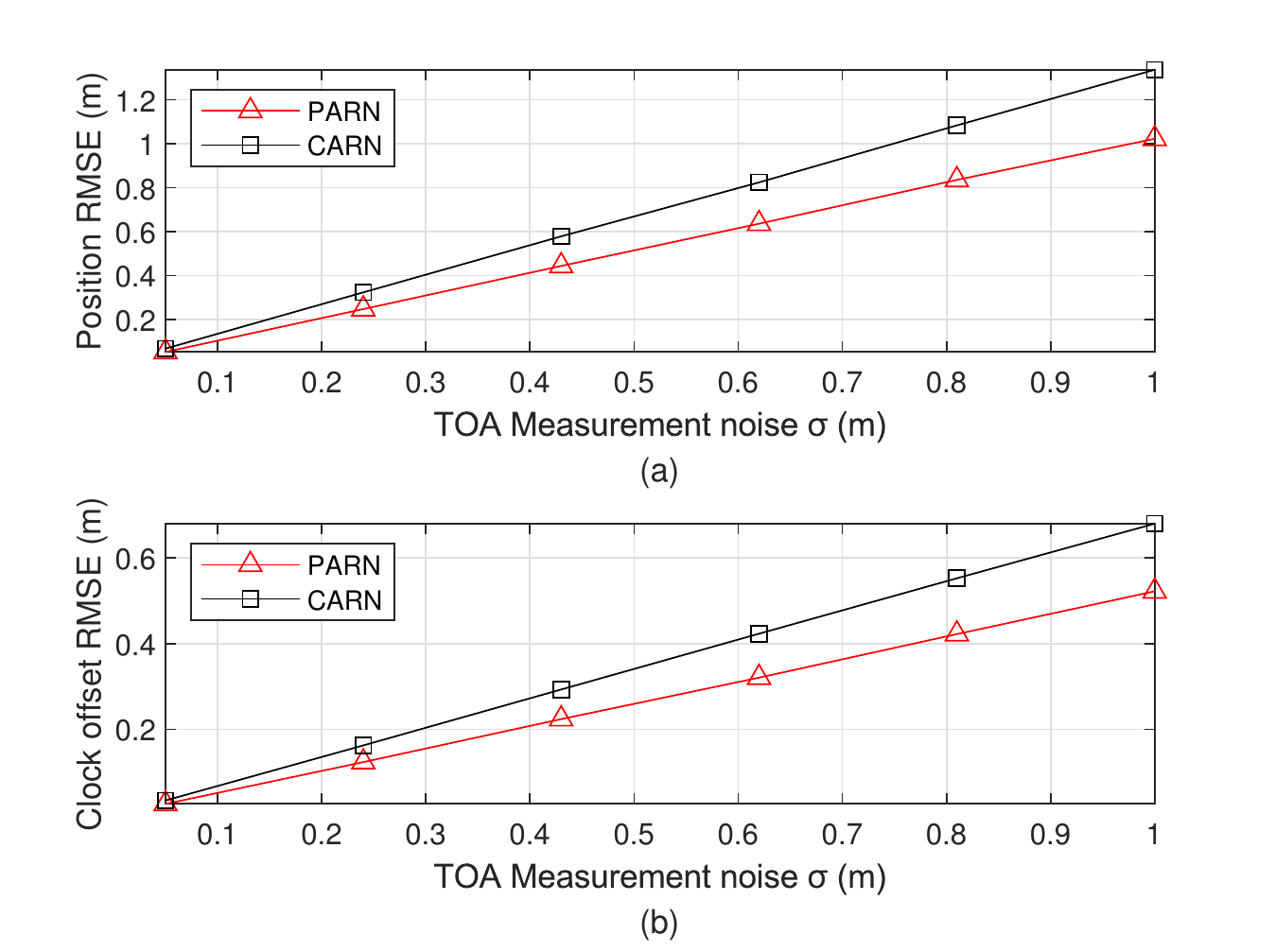} 
	\caption{Localization and synchronization comparison between PARN and the conventional asymmetric ranging network (CARN). The position and clock offset estimation accuracy of the proposed PARN is higher than that of the CARN.
	}
	\label{fig:compareold}
\end{figure}

We also compare the performance of the proposed PARN with that of the conventional asymmetric ranging network, namely CARN, which only uses one-time communication between the PAN and SANs, such as \cite{wang2011robust,amiri2019efficient}. We use the ML-LAS Mode 2 to process the measurements of both systems for fair comparison. We can obtain similar results if we use Mode 1. The localization and synchronization results are shown in Fig. \ref{fig:compareold}. We can see that the proposed PARN outperforms the CARN in both position and clock offset estimation accuracy.

\subsection{Performance of ML-LAS Mode 1 with Deviated UD Velocity} \label{perfromancedV}
We evaluate the localization and synchronization error of Mode 1 (with sync-TOA $\tau_u$) of the proposed ML-LAS method, in the case when the UD velocity information deviates from the true value. In the simulation, we vary the norm of the UD velocity deviation, i.e., $\Vert\Delta \boldsymbol{v}_u\Vert$, from 0 m/s to 20 m/s with a step of 4 m/s. At each step, given $\Vert\Delta \boldsymbol{v}_u\Vert$, the velocity input $\tilde{\boldsymbol{v}}_u$ to the ML-LAS method is randomly selected from those that satisfy $\Vert\Delta \boldsymbol{v}_u\Vert=\Vert\tilde{\boldsymbol{v}}_u-\boldsymbol{v}_u\Vert$. The TOA measurement noise $c\sigma_i$ is set to 0.05 m. We simulate 10,000 samples at each step. Different response delays are set to $\delta t=1$, $5$, $10$ and $25$ ms for comparison.

\begin{figure}
	\centering
	\includegraphics[width=0.99\linewidth]{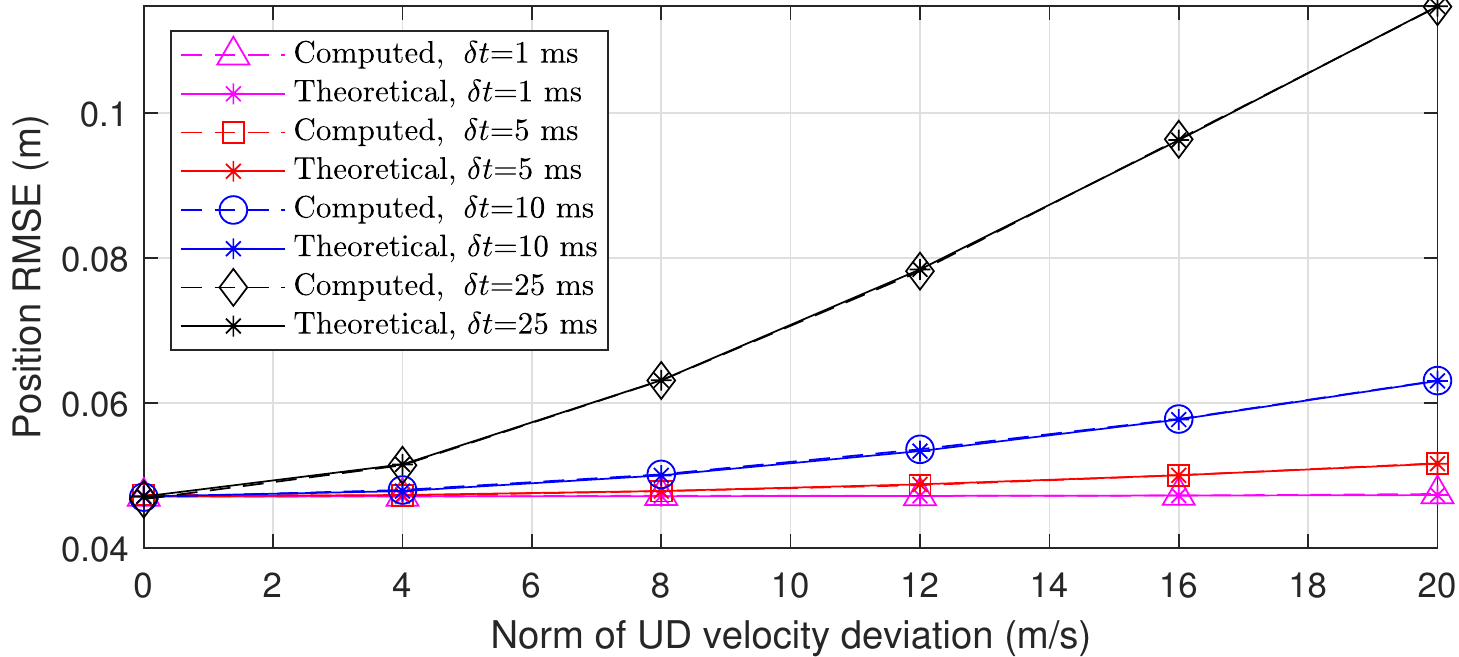} 
	\caption{Position error vs. norm of the UD velocity deviation for ML-LAS Mode 1. The computed position error results match the theoretical analysis. With larger velocity deviation and longer response delay time, the position error increases.
	}
	\label{fig:perrorvsdVnorm}
\end{figure}

\begin{figure}
	\centering
	\includegraphics[width=0.99\linewidth]{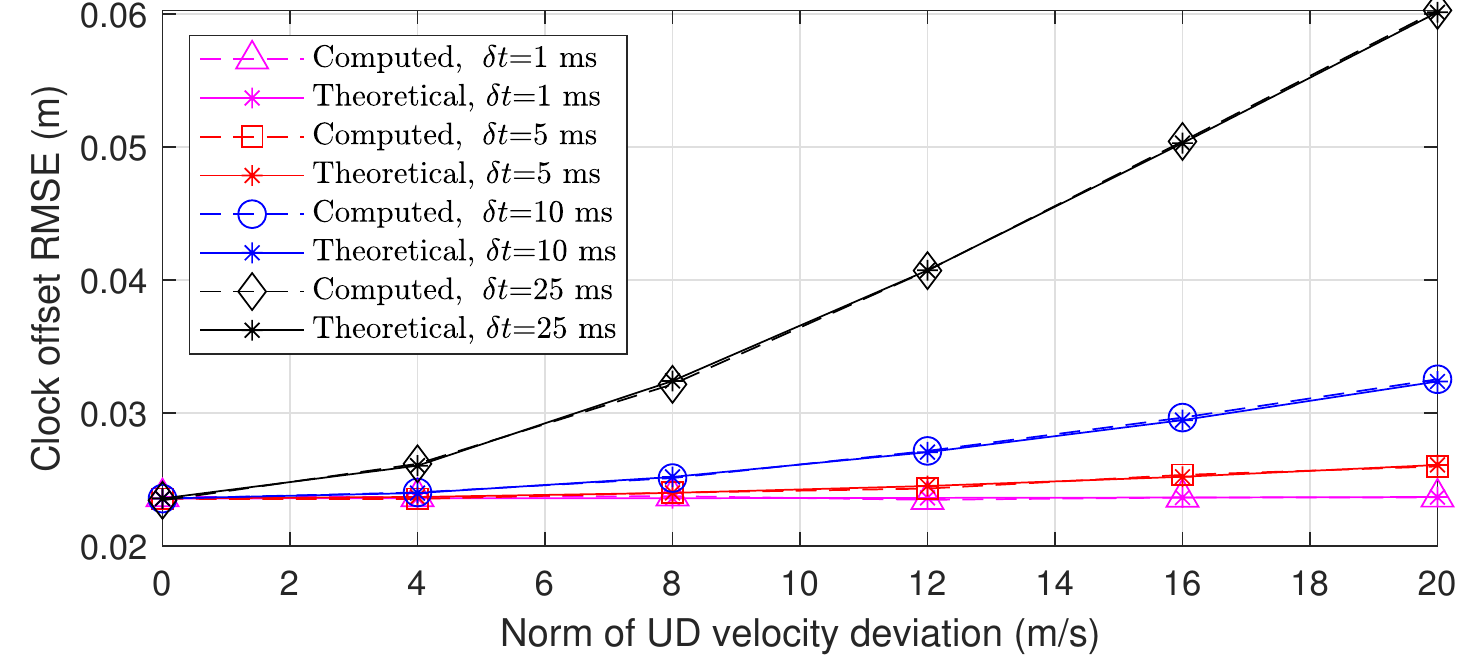} 
	\caption{Clock offset error vs. norm of the UD velocity deviation  for ML-LAS Mode 1. The computed clock offset errors of the proposed ML-LAS method match the theoretical analysis. The synchronization error increases with growing velocity deviation and delay time.
	}
	\label{fig:berrorvsdV}
\end{figure}

The position and the clock offset error results are given by Fig. \ref{fig:perrorvsdVnorm} and Fig. \ref{fig:berrorvsdV}, respectively. Take the error curves of the case with $\delta t=25$ ms as an example. We observe that the estimation errors increase with larger UD velocity deviation. The theoretical curves are generated based on (\ref{eq:mudvsquare1}) and (\ref{eq:RMSEdV}). We can see that the computed RMSEs match the theoretical value. It is also shown that when the velocity deviation is small, the linear growing relation is not obvious due to the measurement noise-caused error. However, when the velocity deviation is large enough, the total error is dominated by the deviation-caused term, and the curves appear more linear. This result is consistent with the theoretical analysis presented in Section \ref{ErrorAnalysis}.

Fig. \ref{fig:perrorvsdVnorm} and Fig. \ref{fig:berrorvsdV} also show that with larger response delay time $\delta t$, the velocity deviation will cause greater localization and synchronization errors. This result offers a guidance for real systems that the response delay needs to be designed as short as possible to reduce the velocity deviation-caused errors.

\subsection{Performance of ML-LAS Mode 1 with Deviated UD Clock Drift}\label{performancedome}
If the input UD clock drift to the proposed ML-LAS method deviates from its true value, the localization and synchronization results will have large errors. In this subsection, we vary the input UD clock drift $\tilde{\omega}_u$, and set the deviation range from 0 ppm to 0.5 ppm, equivalent to the LOS speed ranging from 0 m/s to 150 m/s in a radio frequency (RF) localization system. We simulate 10,000 samples at each of the 6 steps. Different response delays are set to $\delta t=1$, $5$, $10$ and $25$ ms for comparison.

\begin{figure}
	\centering
	\includegraphics[width=0.99\linewidth]{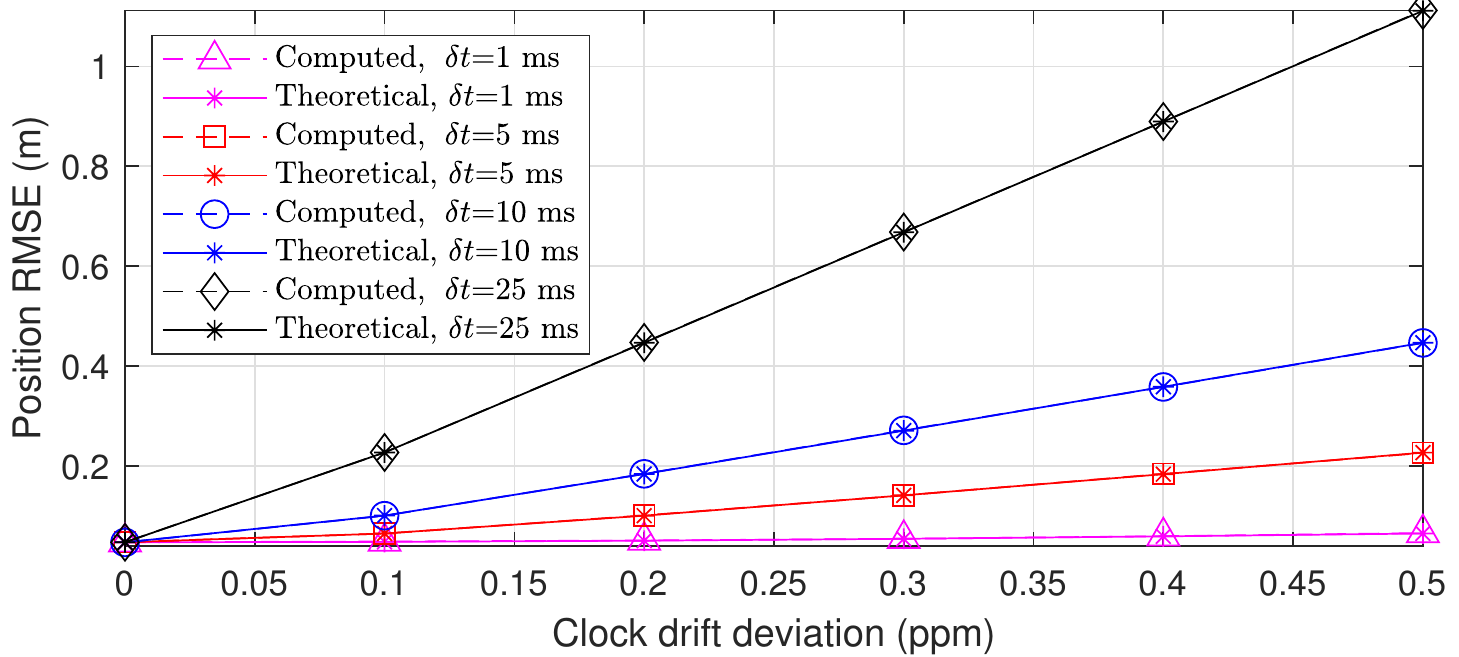}\caption{Position error vs. UD clock offset input deviation for ML-LAS Mode 1. The computed position errors matches the theoretical analysis. The localization estimation error increases with growing input clock drift deviation and delay time.}
	\label{fig:perrordome}
\end{figure}

\begin{figure}
	\centering
	\includegraphics[width=0.99\linewidth]{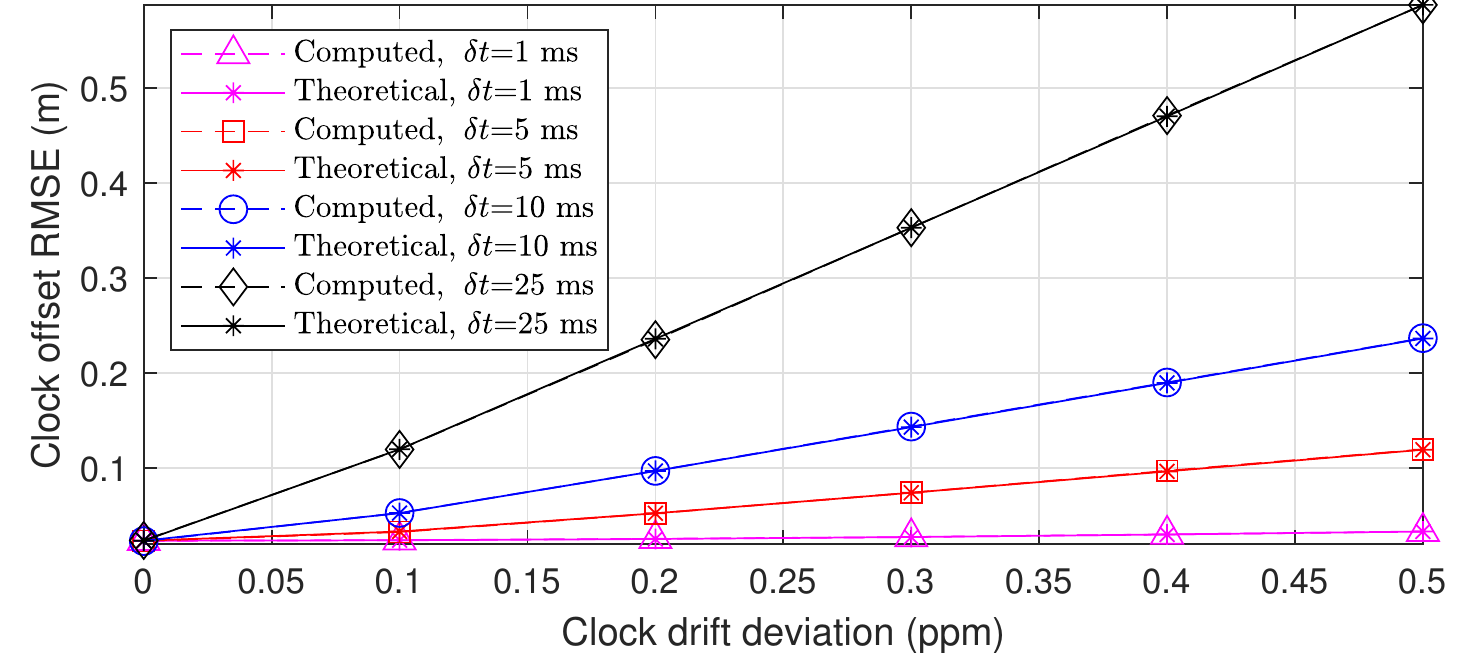} 
	\caption{Clock offset error vs. UD clock offset input deviation for ML-LAS Mode 1. The clock offset estimation errors matches the theoretical analysis. It increases with growing clock drift deviation and delay time.}
	\label{fig:berrordome}
\end{figure}

The position and clock offset error results versus the deviation of the input UD clock drift with different delay times are shown in Fig. \ref{fig:perrordome} and Fig. \ref{fig:berrordome}, respectively. We observe from the figures that the position and clock offset errors become larger when the deviation of the input clock drift grows. The estimation errors match the theoretical values computed by (\ref{eq:RMSEdome}), validating the theoretical analysis in Section \ref{deviateOme}. With shorter response delay $\delta t$, the clock drift deviation-caused errors will decrease. Similar to the results from deviated UD velocity in Section \ref{perfromancedV}, this also shows the necessity of a short response delay in the PARN design to reduce the localization and synchronization errors.

In real applications such as drone control and navigation, the UD clock drift can be measured beforehand and stored for use in the mission. During the flight mission, the clock drift can be updated when the UD is stationary. In the experiment section, we will demonstrate the feasibility of measuring the clock drift using a short data-set collected from a consumer level UWB localization device with a low cost oscillator.

%
%

\section{Real System Experimental Results}\label{experiment}
\subsection{System Setup}
\begin{figure}
	\centering
	\includegraphics[width=0.99\linewidth]{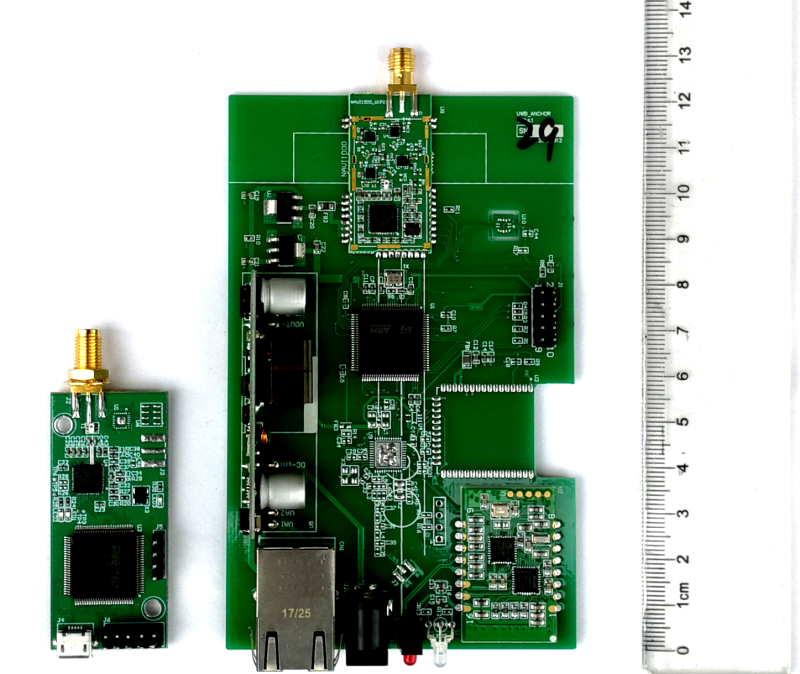} 
	\caption{Circuit boards of prototype UD and AN. Left: UD. Right: AN}
	\label{fig:PCB}
\end{figure}

We have developed a prototype PARN system based on off-the-shelf UWB chips \cite{dw1000ds} and consumer-level micro-controllers. The printed circuit board of the UD and AN are shown in Fig. \ref{fig:PCB}. Experiments are conducted to verify the performance of the proposed new PARN in AN virtual synchronization and UD localization and synchronization.

\begin{figure}
	\centering
	\includegraphics[width=0.99\linewidth]{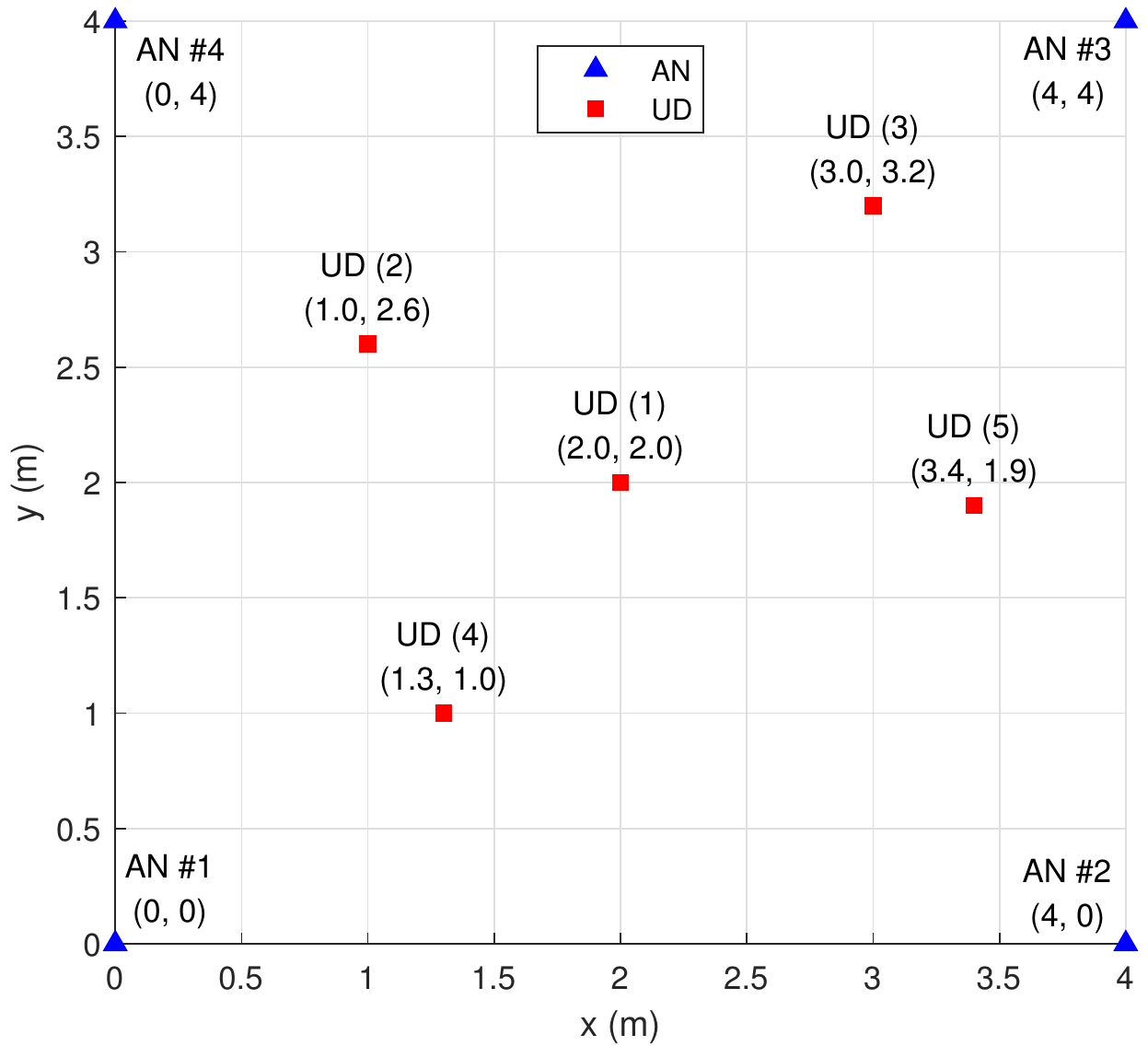} 
	\caption{AN and UD placement for experiment using real prototype system.}
	\label{fig:experimentsetup}
\end{figure}

We place four ANs at the corners of a 4 m $\times$ 4 m area as shown in Fig. \ref{fig:experimentsetup}. AN \#1 is set as the PAN, and transmits sync signal periodically with an interval of 10 ms. The UD transmits response signal after approximately 1 ms delay from the reception of the sync signal. That enables a 100 Hz localization frequency for each UD. The accurate response delay time is measured by the UD. All the transmission and reception timestamps are recorded to form TOA measurements. The UD is placed at five different locations as shown in the same figure. We use a tapeline to measure the position of the UD and ANs, and the accuracy is only better than 5 cm. The TOA measurement noise $c\sigma_b$ and $c\sigma_i$ are both set to 0.03 m. The clock parameters are set to $s_b=10^{-21}$ s and $s_{\omega}=5.9\times 10 ^{-23}$ s$^{-1}$. We collect 10-min TOA measurement data to evaluate the performance of the new system with the proposed Kalman filter-based virtual synchronization and the ML-LAS method.

\subsection{SAN Virtual Synchronization Result}\label{testKalman}
We use the Kalman filter-based method presented in Section \ref{virtualsync} to virtually synchronize the SANs with AN \#1. Unlike in the numerical simulation, the true clock offset of the SAN in the experiment is not attainable. However, we can compare the filtered clock offset with the original TOA measurement to see if the noise is reduced. We note that the clock offset increases or reduces over time and the range of variation during a 10-min time length is much larger than the noise. This makes the noise invisible if we plot the original measurement-versus-time curve in a figure. We thereby apply a difference operation to adjacent measurements (and to the adjacent estimated clock offsets). Then, we divide the differenced value by the time interval as normalization to deal with the case that measurements are lost at some instants. This operation is expressed by
\begin{align}\label{eq:differential}
\text{Differential TOA measurement}=
\frac{\tau_i(n)-\tau_i(n-1)}{t_{TX}^{(1)}(n)-t_{TX}^{(1)}(n-1)} \text{,} \nonumber\\
\text{Differential estimated clock offset}=
\frac{\check b_i(n)-\check b_i(n-1)}{t_{TX}^{(1)}(n)-t_{TX}^{(1)}(n-1)} \text{.}
\end{align}

\begin{figure}
	\centering
	\includegraphics[width=0.99\linewidth]{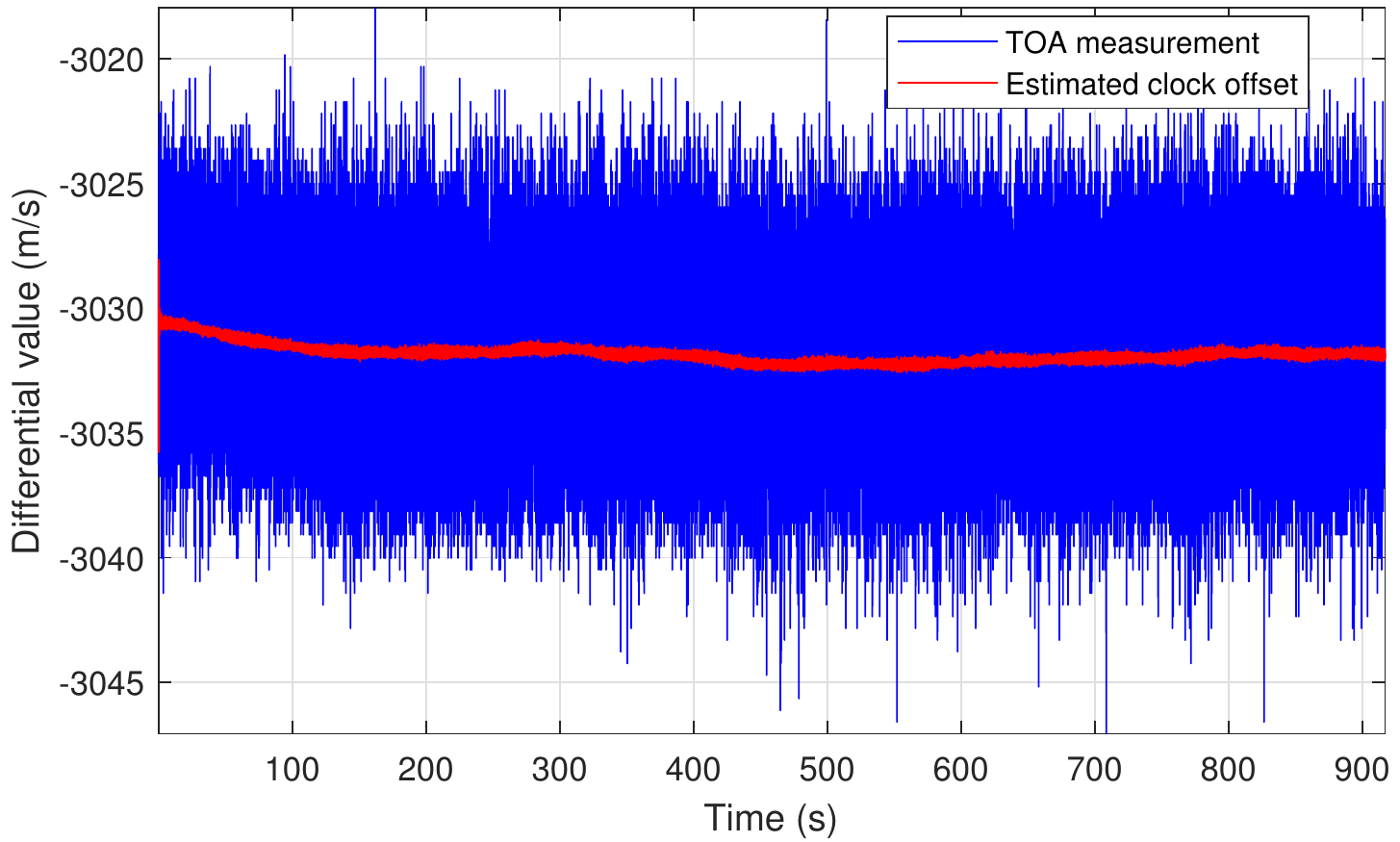} 
	\caption{Differential value comparison between TOA measurements and estimated clock offset from real system experiment. The differential values are obtained from (\ref{eq:differential}). The calculated value is multiplied by the signal propagation speed $c$ for better illustration. The estimated clock offset is from the Kalman filter-based virtual synchronization. The noise is effectively reduced compared with the original TOA measurements.}
	\label{fig:expKFresult}
\end{figure}

After the differential operation, we plot the 15-min results of AN \#2 in Fig. \ref{fig:expKFresult}. We convert the differential value of the clock offset to the speed of displacement, which is commonly adopted for illustration \cite{misra2006global}. The estimated clock offset result from the Kalman filter shows a reduced noise compared with the original TOA measurement. This result validates the effectiveness of the proposed Kalman filter-based method in virtual synchronization for the SANs.

\subsection{UD Localization and Synchronization Result}
We use the two modes (Mode 1 with sync-TOA and Mode 2 without) of the proposed ML-LAS method to produce the localization results for all the five UD locations shown in Fig. \ref{fig:experimentsetup}. The standard deviation (STD) of the localization results from the two modes are listed in Table. \ref{table_resultstd}. Due to limited coordinate measurement accuracy by tapeline, the absolute localization results from the two modes are not helpful in performance evaluation. However, the STD shows their performance difference. As given by Table. \ref{table_resultstd}, we can see that the STD of the localization result from ML-LAS Mode 1 is smaller than that from Mode 2 for all experiments at the five UD locations. This verifies the theoretical analysis given by Section \ref{CRLBanalysis}, and validates the performance of the proposed ML-LAS method in the real world.

The STD of the localization results from the conventional asymmetric ranging network (CARN), which only uses one-time communication between ANs for synchronization, is listed in the last two columns of Table. \ref{table_resultstd}. For fair comparison, we use the ML-LAS Mode 2 to generate the localization results. We can see that the STD of the CARN is larger than that of the PARN. This shows the superior performance of the PARN over the CARN.

\begin{table}[!t]
	\begin{threeparttable}
	\caption{Localization STD for 5 UD locations from real system experiment}
	\label{table_resultstd}
	\centering
	\begin{tabular}{c c c c c c c}
		\toprule
		\multirow{3}{1cm}{\hspace{0.15cm} UD location} 
		&\multicolumn{4}{c}{PARN}&\multicolumn{2}{c}{CARN}\\
		&\multicolumn{2}{c}{ML-LAS Mode 1} &\multicolumn{2}{c}{ML-LAS Mode 2}&\multicolumn{2}{c}{ML-LAS Mode 2}\\
		\cline{2-7}
		\multirow{2}{*}{} &$x$ (cm)& $y$ (cm)& $x$ (cm)& $y$ (cm)& $x$ (cm)& $y$ (cm)\\

		\hline
		(1) & 1.72 & 1.68& 1.87& 1.85& 2.64& 2.63\\
		(2)  & 2.06 & 1.46& 2.13& 1.87& 3.03& 2.48\\
		(3) & 1.95 & 1.90& 2.19& 2.31& 3.14& 3.24\\
		(4) & 1.69 & 1.90& 1.80& 2.01& 2.05& 2.43\\
		(5) & 1.66 & 1.87& 2.47& 1.96& 3.36& 2.73\\
		\bottomrule
	\end{tabular}

\begin{tablenotes}[para,flushleft]
Note: All the STDs from ML-LAS Mode 1 are smaller than that from Mode 2. This result validates the performance of the ML-LAS and shows the superior localization accuracy when utilizing the sync-TOA measurement. The last two columns are results from the conventional asymmetric ranging network (CARN), processed by ML-LAS Mode 2. Compared with the results in PARN, the CARN results have larger STD, showing the superiority of the PARN.
\end{tablenotes}
\end{threeparttable}
\end{table}

\begin{figure}
	\centering
	\includegraphics[width=0.99\linewidth]{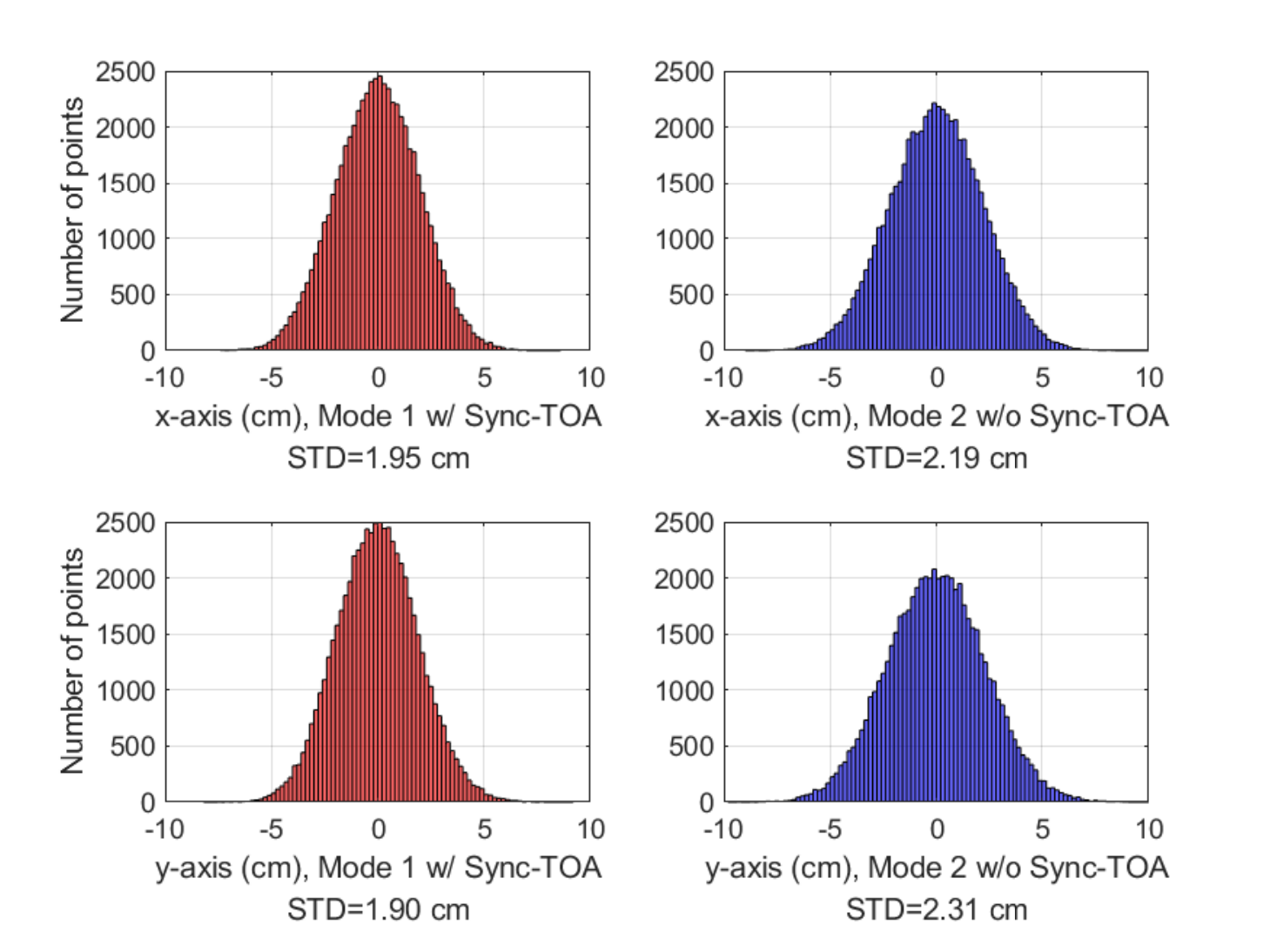} 
	\caption{Histograms of ML-LAS Mode 1 and ML-LAS Mode 2 localization results from real system experiment. The localization results from both modes follow Gaussian distribution. The STD of Mode 1 is smaller than that of Mode 2, demonstrating its better accuracy.}
	\label{fig:exampleresult}
\end{figure}

To see more details in the PARN localization results, we take the data of UD location (3) as an example for investigation. The histograms of the $x$ and $y$-axis localization results are shown in Fig. \ref{fig:exampleresult}. It shows that the localization results from both modes follow Gaussian distribution. The STD of the Mode 1 result is smaller than that of Mode 2 result, indicating that Mode 1 has better localization accuracy than Mode 2. This is consistent with the theoretical analysis and the simulation results given in previous sections.

The STD of the synchronization result or the clock offset estimation from the two modes cannot be obtained directly because the clock offset is difficult to de-trend due to its non-linear change with time. However, given that the UD and PAN are stationary in the experiment, the TOA measurement $\rho_1$ has exactly the opposite trend to the UD clock offset according to (\ref{eq:rhoANiE}), and can be used to de-trend the UD clock offset. Based on (\ref{eq:rhoANiE}), we have
\begin{align} \label{eq:clockbdetrend}
b_{\text{detrend}}=c\rho_1 +c\hat{b}_u-\left\Vert\boldsymbol{p}_1-\boldsymbol{p}_u\right\Vert=  -c\Delta b_u + c\varepsilon_{i} \text{,}
\end{align}
where $b_{\text{detrend}}$ represents the de-trended clock offset, $\hat{b}_u$ is the estimated clock offset by either Mode 1 or Mode 2, and $\Delta b_u$ is the estimation error.

\begin{figure}
	\centering
	\includegraphics[width=0.99\linewidth]{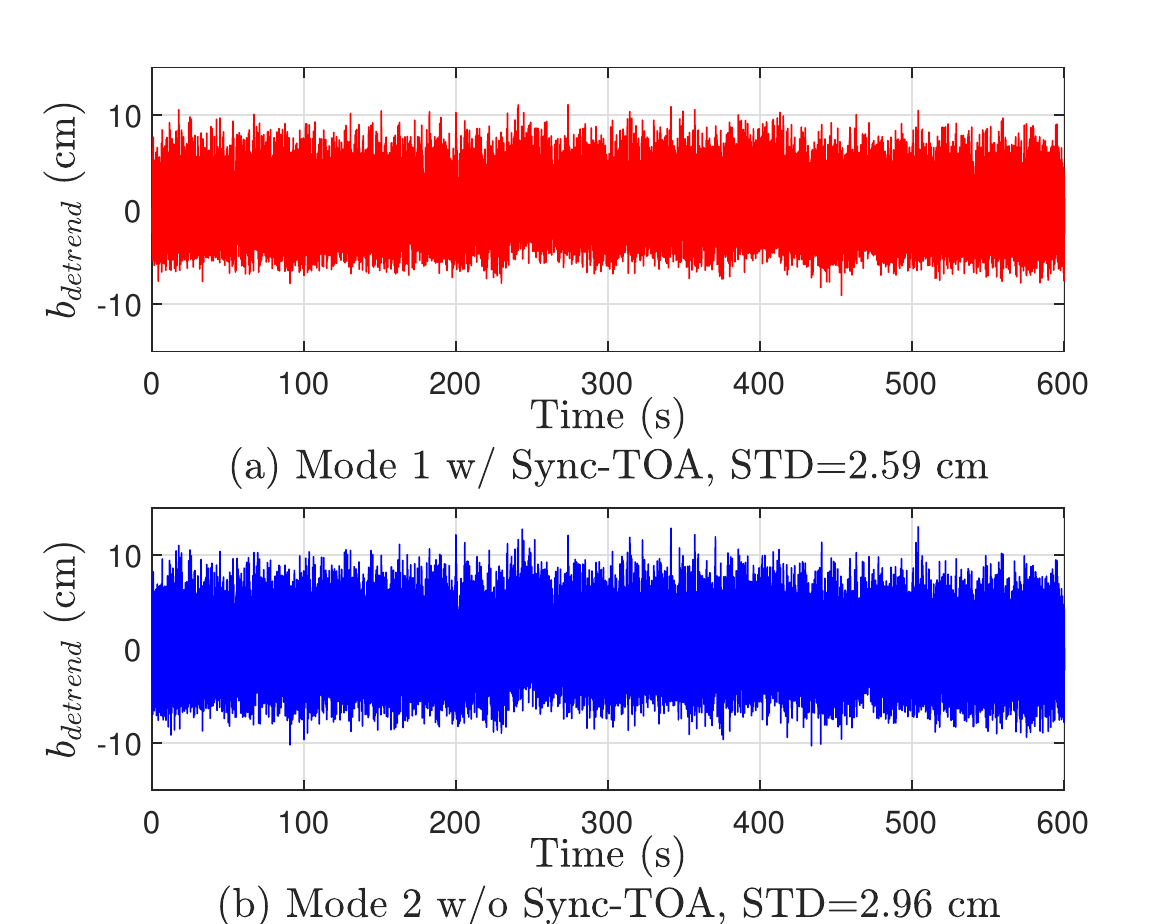}
	\caption{UD clock offset estimation (de-trended by $\rho_1$) vs. time from real experimental data. STD comparison between the two modes shows that Mode 1 of the ML-LAS method yields better synchronization accuracy.}
	\label{fig:clockbvstime}
\end{figure}

We use the data of UD location (3) to analyze the synchronization result. The clock offset estimation results from both modes of the ML-LAS method are added to the TOA measurements $\rho_1$ based on (\ref{eq:clockbdetrend}), and the two obtained curves over time are depicted in Fig. \ref{fig:clockbvstime}. We compute the STDs of the de-trended data from both modes. Note that the de-trended data is a combination of the estimation error $c\Delta b_u$ and the measurement noise $\varepsilon_i$. The measurement noise $\varepsilon_i$ is common for both modes. Thus, the STD is an indicator for the clock offset estimation error. We can see that the results from ML-LAS Mode 1 have smaller STD, indicating a better accuracy in synchronization. This result matches the theoretical analysis and the simulation results in the previous sections.

\subsection{Discussion on UD Clock Drift}
In real-world applications, we have to consider how we obtain the UD clock drift and how accurate it can be so that we can apply Mode 1 of the ML-LAS method. In this subsection, we analyze the real data from our prototype system to show a feasible way to estimate the UD clock drift for use in the proposed ML-LAS method.

We first place the UD and the PAN at two ends of a fixed distance and collect 20-min sync-TOA measurement data. Both their clock sources are a low cost crystal resonator \cite{DSX321Gds}. According to the clock model given by (\ref{eq:clockbomega}), we roughly estimate the UD clock drift by dividing the differenced value of adjacent TOA measurements by the transmission interval, i.e., 
\begin{align}\label{eq:estome}
\text{Rough Estimation} \;\omega_u\approx \frac{\tau_u(n)-\tau_u(n-1)}{t_{TX}^{(1)}(n)-t_{TX}^{(1)}(n-1)} \text{,}
\end{align}

\begin{figure}
	\centering
	\includegraphics[width=0.99\linewidth]{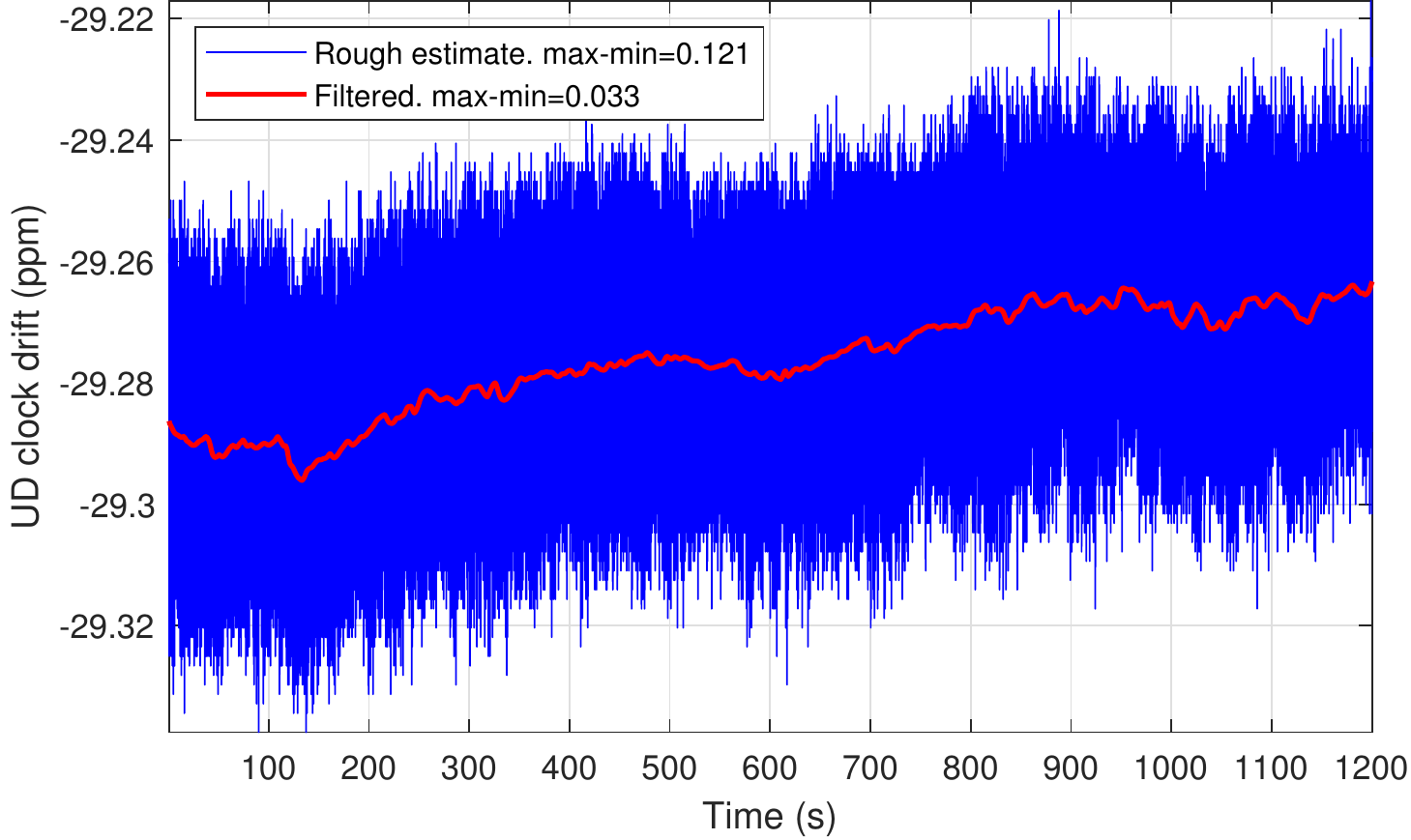}
	\caption{UD clock drift estimation vs. Time from real experimental data. The blue curve is a noisy estimation based on (\ref{eq:estome}). The fluctuation range is 0.121 ppm. The red curve is de-noised by a Kalman filter. The fluctuation range is shrunk down to 0.033 ppm.}
	\label{fig:omegavstime}
\end{figure}

We note that (\ref{eq:estome}) is a very rough estimation of the UD clock drift because the TOA measurements are noisy. The rough estimation of the UD clock drift versus time is plotted as the blue curve in Fig. \ref{fig:omegavstime}. We can see that the trend of the UD clock drift is not a constant and it fluctuates with time. The fluctuation magnitude is smaller than 0.121 ppm, which indicates that the deviation of the clock drift from the true value does not exceed this range. Recall Section \ref{deviateOme} and Fig. \ref{fig:perrordome}, and given the response delay time of the experiment system as 1 ms, the maximal localization error caused by the possible clock drift deviation is less than 1 cm, which does not affect the positioning accuracy of a typical RF localization system. The red curve in the same figure is the filtered result of the UD clock drift processed by the Kalman filter presented in Section \ref{virtualsync}. The fluctuation range is even smaller, down to 0.033 ppm. Hence the localization error caused by the filtered UD clock drift deviation is smaller than 0.3 mm, which is negligible for most IoT localization applications.

The above analysis shows that even with a short period of stationary TOA measurement data, accurate estimate of the UD clock drift can be obtained to ensure satisfactory performance of Mode 1 of the proposed ML-LAS method. In real applications, we can estimate the UD clock drift when the UD is stationary and then use this estimation in the ML-LAS method for localization and synchronization when the UD moves. For systems with tighter requirements, a better clock source such as a high quality temperature compensated crystal oscillator (TCXO) or oven-controlled crystal oscillator (OCXO) can be used to achieve better performance.

\section{Conclusion}
In this article, a new localization and synchronization system based on a periodic asymmetric ranging network, namely PARN, is developed. In this system, the periodic transmission of the PAN sync signal enables high accuracy virtual synchronization between ANs. Utilization of various TOA measurements along with UD velocity and clock drift compensation ensures consistent and simultaneous localization and synchronization for a moving UD. We propose a Kalman filter-based method, which utilizes all historical sync-TOA measurements at the SANs, to achieve high accuracy wireless virtual synchronization for the anchor network. It has low complexity and is suitable for IoT devices. We then propose an optimal ML estimator, namely the ML-LAS method, which employs the response-TOA measurements, the virtual synchronization results for the SANs, and optionally the sync-TOA measurement at the UD end, to simultaneously localize and synchronize a moving UD. We analyze its localization and synchronization error and derive the CRLB. The estimation error caused by deviated UD velocity and clock drift is analyzed quantitatively.


Monte-Carlo simulations are conducted to evaluate the performance of the new PARN. Results show that with the proposed Kalman filter, the PARN can correctly track the clock offset with improved accuracy compared with the original measurements, benefiting the localization and synchronization for the UD. Results also show the optimal performance of the PARN in simultaneous localization and synchronization for a moving UD using the proposed ML-LAS method. Compared with the CARN, which only uses one-time communication between ANs for synchronization, the PARN has higher localization and synchronization accuracy for the UD. The impact of deviated UD velocity and clock drift on the localization and synchronization error is also verified. All numerical results are consistent with the theoretical analysis. We implement a prototype PARN system based on consumer level embedded hardware. Real-world experiment using the prototype system is carried out. Experimental results validate the performance of the Kalman filter-based virtual synchronization for SANs, and the ML-LAS method in UD localization and synchronization. Results demonstrate the feasibility and superiority of the new PARN in the real-world.

\appendices

\section{Kalman Filter Process}
\label{KFprocess}
The Kalman filter process following the notations in Section \ref{virtualsync} is given as follows.

State prediction:
\begin{flalign} \label{eq:statepred}
\check{\boldsymbol{x}}_{n|n-1}=\bm{\Phi}\check{\boldsymbol{x}}_{n} \text{.}
\end{flalign}

Prediction error variance:
\begin{flalign} \label{eq:errorpred}
\bm{P}_{n|n-1}=\bm{\Phi}\bm{P}_{n-1}\bm{\Phi}^T+\bm{Q} \text{.}
\end{flalign}

Kalman gain:
\begin{flalign} \label{eq:Kalmangain}
\bm{K}_{n}=\bm{P}_{n|n-1}\bm{H}^T\left(\bm{H}\bm{P}_{n|n-1}\bm{H}^T+\sigma_i^2\right)^{-1} \text{.}
\end{flalign}

Estimation error variance:
\begin{flalign} \label{eq:errorest}
\bm{P}_{n}=\left(\bm{I}-\bm{K}_{n}\bm{H}\right)\bm{P}_{n|n-1} \text{.}
\end{flalign}

State estimation:
\begin{flalign} \label{eq:stateest}
\check{\boldsymbol{x}}_{n}=\check{\boldsymbol{x}}_{n|n-1}+\bm{K}_n\left(z(n)-\bm{H}\check{\boldsymbol{x}}_{n|n-1}\right) \text{.}
\end{flalign}


%


\ifCLASSOPTIONcaptionsoff
    \newpage
\fi



\bibliographystyle{IEEEtran}
\bibliography{IEEEabrv,paper}
\end{document}